\documentclass[]{jfm}

\usepackage{graphicx}
\usepackage{newtxtext}
\usepackage{newtxmath}
\usepackage{natbib}
\usepackage{hyperref}   
\hypersetup{ 
    colorlinks = true,
    urlcolor   = blue,
    citecolor  = black,
}

\newcommand{\RomanNumeralCaps}[1]  
\linenumbers

\usepackage{epstopdf, epsfig}

\usepackage[hyperref,dvipsnames]{xcolor}
\usepackage{packages/soul/soul}
 \usepackage{packages/bbding/bbding}
 \usepackage{setspace}
 \usepackage{multicol}
 \usepackage{amsmath}
 \usepackage{amssymb}
 \usepackage{mathrsfs}
 \usepackage{amsbsy}
 \usepackage{empheq}
 \usepackage{float}

\usepackage[off,latex={-interaction=nonstopmode},crop=on,runs=2]{auto-pst-pdf}
\usepackage{psfrag}
\usepackage{dashrule}

\usepackage[percent]{overpic}
\usepackage{afterpage}
\usepackage{subcaption}

\usepackage[printfigures]{figcaps}                                             
\figcapsoff 

\newcommand\matlab[1]{}
\definecolor{Green}{RGB}{16,131,16} %

\newcommand\pfrac[2]{\frac{\p #1}{\p #2}}

\newcommand\ts[1]{\textstyle{#1}}

\renewcommand\l{\ell}
\newcommand\N{\mathscr{N}}
\newcommand\M{\mathscr{M}}
\newcommand\Q{\mathscr{Q}}
\newcommand\D{\mathscr{D}}
\renewcommand\a{\alpha}
 
\usepackage{arydshln}

\sbox{\astrutbox}{\rule[-5pt]{0pt}{20pt}}

\renewcommand\t{\ensuremath{\theta}}

\renewcommand*{\doi}[1]{\href{http://dx.doi.org/#1}{DOI: #1}}

\usepackage{tikz}

\shortauthor{A. A. Gyllenberg and R. Sayag}
\title[Lubricated axisymmetric gravity currents of power-law fluids]{Lubricated
  axisymmetric gravity currents of power-law fluids.}

\author{
  Ayala A. Gyllenberg\aff{1},
  \and Roiy Sayag\aff{2,1,3}
 \corresp{\email{roiy@bgu.ac.il}}
}
 
\affiliation{

  \aff{1}Dept. of Mech. Eng., Ben-Gurion University of the
  Negev, Beer-Sheva 8410501, Israel
  \aff{2}Dept. of Environ. Physics, BIDR, Ben-Gurion University of the Negev,
  Sde Boker 8499000, Israel
  \aff{3}Dept. of Physics, Ben-Gurion University of the
  Negev, Beer-Sheva 8410501, Israel}
\begin{document}

\maketitle

 


\begin{abstract}
  The motion of glaciers over their bedrock or drops of fluid along a
  solid surface can become unstable when these substrates are
  lubricated.  Previous studies modeled such systems as coupled
  gravity currents (GCs) consisting of one fluid that lubricates the
  flow of another fluid, and having two propagating fronts. When both
  fluid are Newtonian and discharged at constant flux, global
  similarity solutions were found. However, when the top fluid is
  strain-rate softening experiments have shown that each fluid front
  evolved with a different exponent.  Here we explore theoretically
  and numerically such lubricated GCs in a model that describes the
  axisymmetric spreading of a power-law fluid on top of a Newtonian
  fluid, where the discharge of both fluids is power law in time. We
  find that the model admits similarity solutions only in specific
  cases, including the purely Newtonian case, for a certain discharge
  exponent, at asymptotic limits of the fluids viscosity ratio, and at
  the vicinity of the fluid fronts. Generally, each fluid front has a
  power-law time evolution with a similar exponent as a non-lubricated
  GC of the corresponding fluid, and intercepts that depend on both
  fluid properties. Consequently, we identify two mechanisms by which
  the inner lubricating fluid front outstrips the outer fluid front.
  Many aspects of our theory are found consistent with recent
  laboratory experiments. Discrepancies suggest that hydrofracturing
  or wall slip may be important near the fronts. This theory may help
  to understand the dynamics of various systems, including surges and
  ice streams.
\end{abstract}


\section{Introduction}

Gravity-driven flows of one fluid over another can involve complex
interactions between the two fluids, which can lead to a rich
dynamical behavior.  Such flows occur in a wide range of natural and
human-made systems, as in lava flow over less viscous lava
\citep{Griffiths:2000-ARFM-dynamics,Balmforth:2000-Visco-plastic},
spreading of the lithosphere over the mid-mantle boundary
\citep{Lister-Kerr-1989:the,DauckBoxGellEtAl:2019--Shock}, 
ice flow over an ocean
\citep{DeContoPollard:2016--Contribution,KivelsonKhuranaRussellEtAl:2000--Galileo}
and over bedrock consisting of sediments and water
\citep[][]{Stokes-Clark-Lian-et-al-2007:ice,Fowler:1987-theory}, flows
in permeable rocks \citep{WoodsMason:2000-JFM-dynamics}, liquid drops
deforming over lubricated surfaces
\citep{DanielTimonenLiEtAl:2017-NP-Oleoplaning}, and droplets motion
on liquid-infused surfaces
\citep{KeiserKeiserClanetEtAl:2017-SM-Drop}.

The flow of GCs in circular geometry has been studied
with a range of boundary conditions. In the absence of a lubricating
layer, a common boundary condition along the base of a sole GC is no
slip. Such GCs of Newtonian fluids that are discharged at a rate
proportional to \(t^\a\), where \(t\) is time and $\a$ is a
non-negative scalar, has a similarity solution in which the front
evolves proportionally to \(t^{(3\a + 1)/8}\)
\citep{Huppert:1982-JFM-Propagation}.  Similar axisymmetric gravity
currents of power-law fluids having exponent $n$, where $n=1$
represents a Newtonian fluid and $n>1$ represents a strain-rate
softening fluid, also have similarity solutions in which the front
propagation is proportional to $t^{[\a(2n+1) +1]/(5n+3)}$
\citep{Sayag-Worster:2013-Axisymmetric}.
  
On the other extreme, the presence of a lower fluid layer can
significantly reduce friction at the base of the top fluid, resulting
in extensionally dominated GCs. This is the case, for example, for ice
shelves, which deform over the relatively inviscid oceans with weak
friction along their interface.  The late-time front evolution of such
axisymmetric GCs of Newtonian fluids is proportional to $t$ and is
believed to be stable
\citep{PeglerWorster:2012-JFM-Dynamics}. However, when the top fluid
is strain-rate softening, an initially axisymmetric front can
destabilise and develop fingering patterns that consist of tongues
separated by rifts \citep{SayagWorster:2019-JFM-Instability1}.
         
 
In the more general case friction along the base of GCs can vary
spatiotemporally, as their stress field evolves. For example, the
interface of an ice sheet with its underlying bed rock can include
distributed melt water and sediments, which impose nonuniform and
time-dependent friction along the ice base, and evolve
spatiotemporally under the stresses imposed by the ice layer
{\citep{SchoofHewitt:2013-ARFM-Ice,Fowler:1981-PTotRSoLSAMaPS-theoretical}}.
Consequently, the coupled ice-lubricant system may evolve various flow
patterns {such as ice streams
  \citep[][]{Stokes-Clark-Lian-et-al-2007:ice,Kyrke-Smith:2013-Subglacial}
  and glacier surges \citep[][]{Fowler:1987-theory}}.
Such flows were also modelled as two coupled GCs of
Newtonian fluids spreading axisymmetrically one on top of the other
\citep{KowalWorster:2015-JFM-Lubricated}.  The early stage of these
flows follows a self-similar evolution, in which the fronts of the two
fluids evolve like $t^{1/2}$, as in non-lubricated (no-slip) GCs 
\citep{Huppert:1982-JFM-Propagation}, but they can have a radially
non-monotonic thickness.
Furthermore, laboratory experiments were found to be consistent with
the similarity solutions after an initial transient state, but at a
later stage they became unstable and developed fingering patterns
\citep{KowalWorster:2015-JFM-Lubricated}.
It has been suggested that such instabilities appear when the jump in
hydrostatic pressure gradient across the lubrication front is negative
\citep{KowalWorster:2019-JFM-Stability,KowalWorster:2019-JFM-Stabilityb}.

\begin{figure}
  \hspace{-7mm}
  \includegraphics[scale=0.5]{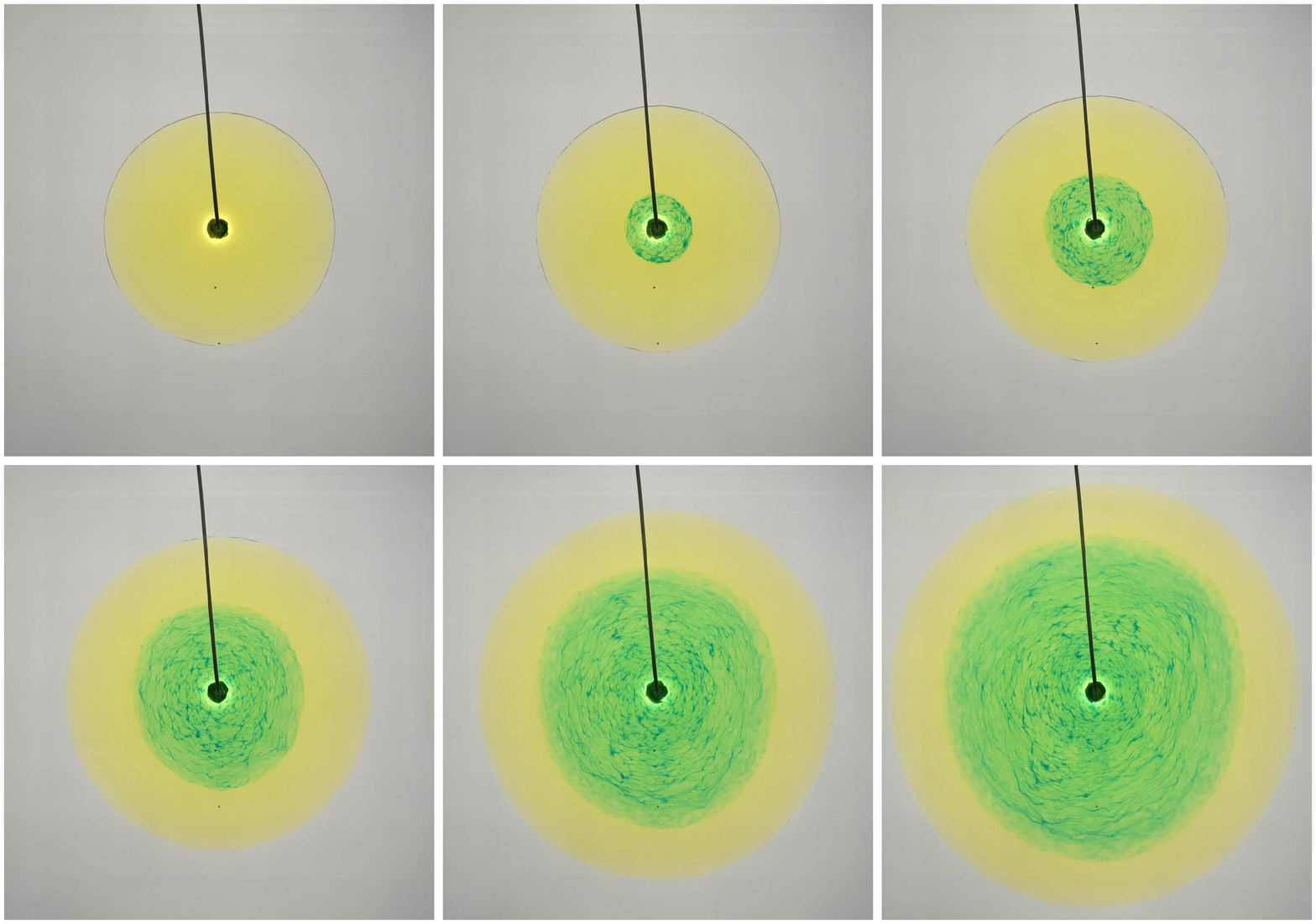}
  \begin{picture}(0,0)(0,0)
    \put(13,183){$t/t_L=1$}
    \put(135,183){$t/t_L=1.13$}
    \put(259,183){$t/t_L=1.33$}
    \put(13,52){$t/t_L=1.85$}
    \put(135,52){$t/t_L=2.64$}
    \put(259,52){$t/t_L=3.57$}
  \end{picture}
\vspace*{-15mm}
\caption{Snapshots from a laboratory experiment
  \citep{KumarZuriKoganEtAl2021JoFMLubricated} of a lubricated GC
  that consists of a strain-rate softening fluid (yellow) lubricated
  by a sugar solution (blue, appears green).  Marked time at each
  snapshot is relative to the
  initiation time $t_L$ of the lubricating fluid.
  \label{fig:Kumar}}
\end{figure}

Despite the wide range of natural lubricated GCs that involve
non-Newtonian fluids, the radial flow of a non-Newtonian fluid over a
lubricated layer of Newtonian fluid has just recently been explored
experimentally \citep{KumarZuriKoganEtAl2021JoFMLubricated}. Motivated
by glacier flow over lubricated bedrock, the experimental setup
consisted of a GC of a strain-rate-softening fluid
(Xanthan-gum solution) that has a power-law viscous deformation
similar to ice \citep{Glen-1952:experiments}, and a lubricating
GC of a less viscous Newtonian fluid (diluted sugar
solution).
The pattern of both fluids in those constant-flux $(\alpha=1)$
experiments remained axisymmetric throughout the flow (Figure
\ref{fig:Kumar}), in contrast to the fingering pattern that emerged in the
purely Newtonian experiments \citep{KowalWorster:2015-JFM-Lubricated},
as long as the flux ratio of the lubricating fluid to the
non-Newtonian fluid was lower than $\sim 0.06$.
The fronts of the two fluids appeared to have a power-law time
evolution with different exponents. Particularly, the front of the top
non-Newtonian fluid evolved with the same exponent $(2n+2)/(5n+3)$ as
a non-lubricated power-law fluid
\citep{Sayag-Worster:2013-Axisymmetric}, whereas the front of the
Newtonian lubricating fluid evolved with an exponent $1/2$ similar to
a Newtonian non-lubricated GC
\citep{Huppert:1982-JFM-Propagation}. Despite the similarity of the
exponents, the fronts of those lubricated GCs evolved faster than the
corresponding non-lubricated GCs owing to larger intercepts. In
addition, in contrast with the monotonically declining thickness of
non-lubricated GCs, the thickness of the lubricated, non-Newtonian
fluid was found to be nearly uniform in the lubricated part of the
flow, while that of the lubricating fluid was non-monotonic with
localised spikes.

Following up the experimental study of
\citet{KumarZuriKoganEtAl2021JoFMLubricated}, here we develop the
theory for lubricated axisymmetric GCs of power-law
fluids, and explore its major consequences.  Specifically, we develop
a mathematical model for axisymmetric flow of a viscous gravity
current of a power-law fluid lubricated by a Newtonian gravity
current, considering a general input flux of the form \(t^{\a}\), and
show that in the general case the flow has no global similarity
solution of the 1st kind (\S \ref{sec:model}). We then describe the
numerical solver (\S\ref{sec:numerical}) and investigate several
special cases that have similarity solutions
(\S\ref{sec:similarity}). We then explore the possibility that the
inner lubricating front outstrip the outer front
(\S\ref{sec:outstripping}). Finally, we investigate the case of
constant flux discharge (\S\ref{sec:const}), and compare our
theoretical predictions with the laboratory experiments of
\citet{KumarZuriKoganEtAl2021JoFMLubricated} (\S
\ref{sec:comparison}).

\section{Mathematical Model}
\label{sec:model}

\begin{figure}
\centering
\includegraphics[scale=1]{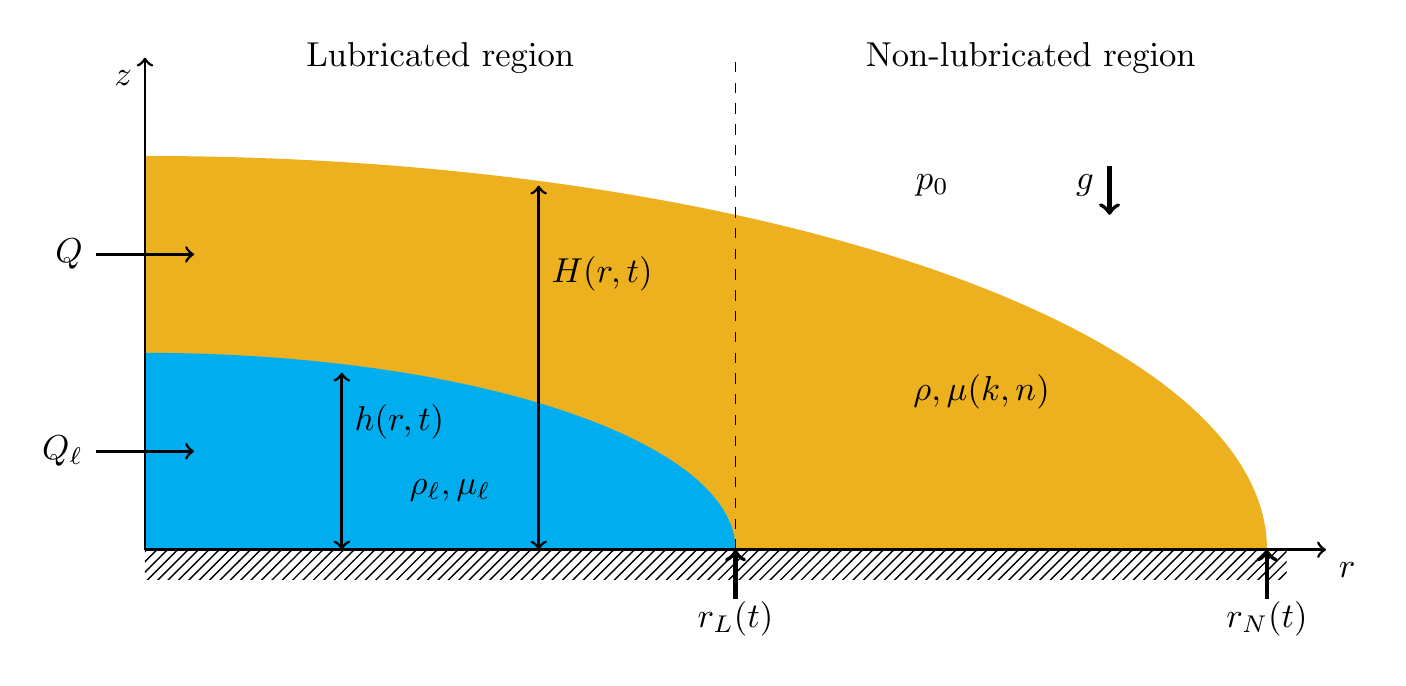}
\vspace*{-7mm}
\caption{Diagram illustrating a GC of Newtonian fluid
  (blue) lubricating a GC of a power-law fluid (yellow).
\label{diagram}} 
\end{figure}

Consider a power-law fluid having viscosity \(\mu\) and density
\(\rho\) that spreads axisymmetrically under its own weight over a
horizontal rigid surface. Simultaneously, a lubricating film of
Newtonian fluid of viscosity \(\mu_\l\) and density \(\rho_\l\) spreads
axisymmetrically over the substrate and below the power-law fluid
(Figure \ref{diagram}). Both fluids are discharged at the origin of a
cylindrical coordinate system in which \(r\) is the radial coordinate
and \(z\) the vertical coordinate. The upper and lower fluid fronts
are denoted by \(r_N(t)\) and \(r_L(t)\) respectively, and the
corresponding fluid thicknesses are denoted by \(H(r,t)-h(r,t)\) and
\(h(r,t)\) respectively, where \(t\) is the time variable. Beginning
the discharge of the lubricating fluid at a time delay \(t_L\) with
respect to the upper fluid, creates two regions in the flow: an inner,
lubricated region (\( r \le r_L\)) in which the power-law fluid flows
at a finite velocity along the interface with the lubricating fluid,
and an outer, non-lubricated region (\(r_L < r \le r_N\)) in which the
power-law fluid meets the substrate and has zero velocity along it.

Assuming that the radial extent of the flow is much greater than its
thickness in both fluid layers, we can apply the lubrication
approximation, which implies that the flow is primarily horizontal (a
GC) and that the dominant strain rate in each fluid layer
$i$ is \(\partial u_i / \partial z\), where \(u_i(r,z,t)\) is the
radial velocity component. Therefore, the axisymmetric Cauchy
equations for each fluid layer, simplified for flows of low Reynolds
numbers and lubrication approximations are
\begin{subequations} \label{eq:cauchy}
\begin{eqnarray} 
\pfrac{ p_i}{ r} &=&\pfrac{}{ z} \left( \mu_i  \pfrac{ u_i}{ z} \right),\label{eq:4}\\
\pfrac{ p_i}{ z}&=&-\rho_i g, \label{eq:hydro4}\\
  \frac{1}{r}\pfrac{ (ru_i)}{ r}+\pfrac{ w_i}{ z}&=&0, \label{eq:cont4}
\end{eqnarray}
\end{subequations}
to leading order
, where \(w_i\) is the vertical velocity component, \(p_i\) is the
pressure field, and \(g\) is the gravitational acceleration. The
pressure is determined hydrostatically in equation \ref{eq:cauchy}b
because we assume a large Bond number, implying that the effects of
surface tension are negligible compared to gravity. We note that
equations \ref{eq:cauchy} satisfy the lubrication approximation, so
this model does not accurately describe the problem at early times,
before the fluid radial extent is sufficiently greater than its
thickness.

The viscosity of the power-law fluid is determined by
\begin{equation} \label{constitutive}
\mu=k \left(\frac{1}{2} \textbf{e}:\textbf{e} \right)^{\ts{\frac{1}{2}\left({\frac{1}{n}-1}\right)}},
\end{equation}
where \(\textbf{e}\) is the strain-rate tensor, $k$ is the consistency
coefficient, and \(n\) is the power-law exponent, which determines the
fluid's response to stress. Specifically, \(n=1\) represents a
Newtonian fluid with dynamical viscosity $k$, \(n > 1\) a
shear-thinning fluid, and \(n < 1\) a shear-thickening fluid. In the
lubrication limit
\(\textbf{e}:\textbf{e} \approx \left(\partial u / \partial
  z\right)^{2}\) to leading order, so that the viscosity of the
power-law fluid simplifies to
\begin{equation} \label{eq:3}
\mu=k \left| \frac{1}{2}\frac{\partial u}{\partial z}\right|^{\ts{\frac{1}{n}-1}}.
\end{equation}

We assume that the total volumes of the power-law fluid and the
Newtonian fluid evolve following a power law in time given by
\begin{subequations} \label{eq:vol}
\begin{equation} \label{eq:volTop}
2 \pi \int _{0} ^{r_N} (H-h)r ~ dr=Q t^{\alpha},
\end{equation}
\begin{equation} \label{eq:volBot}
2 \pi \int _{0} ^{r_L} hr ~ dr=Q_\l\left(t-t_L\right)^{\alpha},
\end{equation}
\end{subequations}
respectively, where \(\alpha\) is a constant exponent, and \(Q\) and
\(Q_\l\) are constant coefficients representing, for instance, the
discharge flux of each fluid when \(\alpha=1\).

\subsection{The Non-Lubricated Region} \label{non-lubricated} In the
non-lubricated region, the power-law fluid meets the substrate and the
lubricating fluid is absent. Integrating \ref{eq:hydro4}, the pressure
distribution is
\begin{equation} \label{eq:hydro}
p(z,r,t)=p_0+\rho g(H(r,t)-z),
\end{equation} 
where \(p_0\) is the ambient pressure over the top free surface of the
power-law fluid. Integration of the radial force balance (\ref{eq:4})
across the depth of the fluid layer together with no-slip boundary
conditions along the substrate and no stress along the free surface
\begin{equation}
u(z=0)=0, \quad \mu \frac{\partial u}{\partial z}(z=H)=0,
\end{equation}
gives the radial velocity field 
\begin{equation} \label{eq.vel} u(z,r,t)=2^{{{1-n}}}\left(\frac{\rho
      g}{k}\right) ^{{n}} \frac{\partial H}{\partial r} \left|
    \frac{\partial H}{\partial r} \right| ^{{n-1}}
  \frac{1}{1+n} \left[(H-z)^{{n+1}}-H^{{n+1}}\right].
\end{equation}
Similar integration of the continuity equation (\ref{eq:cont4}),
accounting for a free surface at the upper boundary \(z=H(r,t)\), and
assuming no normal flow through the substrate, gives the Reynolds
equation 
\begin{equation} \label{eq.reyNL}
\pfrac{ H}{ t}+\frac{1}{r}\pfrac{(rq)}{ r}=0,
\end{equation}
which should satisfy the boundary conditions
\begin{equation} \label{eq.b1}
H(r=r_N)=0 \quad \textrm{and} \quad q(r=r_N)=0,
\end{equation}
where the local flux \(q\) is determined using eq. \ref{eq.vel} to
give
\begin{equation} \label{eq:qNonLub}
q=\int ^{H} _{0} u~dz=-2^{1-n}\left(\frac{\rho g}{k}\right)^{{n}}\pfrac{ H}{ r}\left|\pfrac{ H}{ r} \right|^{{n-1}}\frac{1}{n+2}H^{{n+2}}.
\end{equation}

\subsection{The Lubricated Region} \label{lubricated}
As in the non-lubricated region and assuming continuity of pressure at
the fluid-fluid interface \(z=h\), the pressure in the lubricated
region is determined hydrostatically by
\begin{subequations}
\begin{eqnarray}
p(z)=p_0+\rho g(H-z),& &\quad \textrm{where} \quad h \leq z \leq H,\\
p_\l (z)=p_0+\rho g(H-h)&+\,\rho _\l g(h-z), &\quad \textrm{where} \quad 0 \leq z \leq h.
\end{eqnarray}
\end{subequations}
The radial force balances simplify to
\begin{subequations} \label{eq.214}
\begin{eqnarray} 
\frac{k}{2^{{\frac{1}{n}-1}}}\frac{\partial}{\partial z}\left( \frac{\partial u}{\partial z}\left| \frac{\partial u}{\partial z}\right| ^{\ts{\frac{1}{n}-1}} \right)&=&\frac{\partial p}{\partial r}, \quad \textrm{where} \quad h \leq z \leq H,\\
\mu _\l \frac{\partial ^2 u_\l}{\partial z^2}&=&\frac{\partial p_\l}{\partial r}, \quad \textrm{where} \quad 0 \leq z \leq h,
\end{eqnarray}
\end{subequations}
together with the  boundary conditions,
\begin{subequations} \label{eq.215}
\begin{eqnarray}
u_\l=0,&& \quad \textrm{where} \quad z=0,\\
  u_\l=u, \qquad k\left|\frac{1}{2} \pfrac{ u}{ z}\right| ^{\ts{\frac{1}{n}-1}}\pfrac{ u}{ z}=\mu _\l \frac{\partial u_\l}{\partial z},&& \quad \textrm{where} \quad z=h,\\
\frac{\partial u}{\partial z}=0,&& \quad \textrm{where} \quad z=H,
\end{eqnarray}
\end{subequations}
which represent, respectively, no-slip along the solid substrate,
continuous flow velocity and shear stress at the fluid-fluid
interface, and no shear stress along the free surface of the power-law
fluid. Integrating the radial force balance (equations \ref{eq.214})
across the thickness of each fluid layer, and using the boundary
conditions (equations \ref{eq.215}) we get the radial velocity fields
in each fluid layer
\begin{subequations} \label{eq.uul}
\begin{eqnarray}\label{eq.u}
  u(z,r,t)&=& 2^{1-n}\left(\frac{\rho g}{k}\right)^{{n}}\frac{1}{1+n} \frac{\partial H}{\partial r} \left| \frac{\partial H}{\partial r}\right|^{n-1} \left[(H-z)^{{1+n}}-(H-h)^{{1+n}}\right]\\&&-\frac{\rho g}{\mu _\l}\left[\frac{\partial H}{\partial r}\left(Hh-\frac{h^2}{2}\right)+\frac{h^2}{2}\frac{\rho _\l -\rho}{\rho}\frac{\partial h}{\partial r}\right],\nonumber \\
  u_\l(z,r,t)&=&-\frac{\rho g}{\mu _\l}\left[\frac{\partial H}{\partial r}\left(Hz-\frac{z^2}{2}\right)+\frac{1}{2}\frac{\rho _\l -\rho}{\rho}\frac{\partial h}{\partial r}\left(2hz-z^2\right)\right]. \label{eq.ul}
\end{eqnarray}
\end{subequations}
The Reynolds equations corresponding to each fluid layer have a
similar form as that of the non-lubricated region
\begin{subequations}
  \label{eq:reyL}
\begin{eqnarray}
\frac{\partial h}{\partial t}+\frac{1}{r}\frac{\partial (rq_\l)}{\partial r}&=&0,\\
\frac{\partial (H-h)}{\partial t}+\frac{1}{r}\frac{\partial (rq)}{\partial r}&=&0,
\end{eqnarray}
\end{subequations} 
and should satisfy the boundary conditions
\begin{subequations} \label{eq.end}
\begin{eqnarray}
h=0, \quad q_\l=0 \quad &&\textrm{at} \quad r=r_L,\\
 \textrm{and }\quad q^+=q^-, \quad H^+=H^- \quad&& \textrm{at} \quad r=r_L,\label{eq.218b}
\end{eqnarray}
\end{subequations}
where equation \ref{eq.218b} signifies flux and height continuity
across \(r_L\). The local fluxes result from integrating equations
\ref{eq.uul} to get
\begin{subequations}
  \label{eq:qLubricated}
\begin{eqnarray}
  q=\int_h ^H u~dz &=&-2^{1-n}\left(\frac{\rho
                       g}{k}\right)^{{n}} \frac{\partial
                       H}{\partial r} \left| \frac{\partial
                       H}{\partial
                       r}\right|^{{n-1}}\frac{1}{n+2}(H-h)^{{n+2}}\nonumber\\
                   &&-(H-h)\frac{h \rho g}{\mu _\l}\left[\frac{\partial
                      H}{\partial r}(H-h)+\frac{h}{2}\left(\frac{\partial
                      H}{\partial r}+\frac{\rho
                      _\l-\rho}{\rho}\frac{\partial h}{\partial
                      r}\right)\right],\label{qL}\\
  q_\l=\int_0 ^h u_\l~dz &=&-\frac{h^2}{2}\frac{\rho g}{\mu
                           _\l}\left[\frac{\partial H}{\partial
                           r}(H-h)+\frac{2h}{3}\left(\frac{\partial
                           H}{\partial r}+\frac{\rho
                           _\l-\rho}{\rho}\frac{\partial h}{\partial
                           r}\right)\right].\label{ql}                         
\end{eqnarray}
\end{subequations}
We note that without a lubricant, this set of PDEs is reduced to the
non-lubricated region, and that this PDE set is consistent with the
non-lubricated power-law GC model
\citep{Sayag-Worster:2013-Axisymmetric}, which converges to the
Newtonian non-lubricated GC when \(n=1\) and the fluid
dynamic viscosity is \(k=\mu\)
\citep{Huppert:1982-JFM-Propagation}. In addition, the full PDE set,
for the Newtonian case (\(n=1\)) and constant source fluxes
(\(\alpha=1\)) is consistent with the model for lubricated, Newtonian
GCs \citep{KowalWorster:2015-JFM-Lubricated}.

\subsection{Dimensionless Equations} \label{dimensionless} We
non-dimensionalize equations \ref{eq:vol},
\ref{eq.reyNL}-\ref{eq:qNonLub} and \ref{eq:reyL}-\ref{eq:qLubricated}
with the following time, length and height scales
\begin{subequations}
  \label{eq:scales}
  \begin{eqnarray}
    t &\equiv &T \hat{t},\\
  r &\equiv& \left( \left(\frac{\rho g}{k}\right)^n Q^{2n+1}  T^{\a(1+2n)+1} \right)^{\ts{\frac{1}{5n+3}}} \hat{r},\\
(H,h)&\equiv& \left( \left(\frac{\rho g}{k}\right)^{-2n} Q^{1+n} T^{\a(1+n)-2}  \right)^{\ts{\frac{1}{5n+3}}} (\hat{H},\hat{h}),
\end{eqnarray}
\end{subequations}
where hats denote dimensionless quantities. The resulted dimensionless model, dropping hats, in the non-lubricated region \(r_L \leq r \leq r_N \) is
\begin{equation} \label{eq.ndimb}
\frac{\partial H}{\partial t}+\frac{1}{r}\frac{\partial (rq)}{\partial r}=0,
\end{equation}
where
\begin{equation} \label{eq.q_noslip}
q=-\mathscr{N}\frac{\partial H}{\partial r}\left|\frac{\partial
    H}{\partial r} \right|^{{n-1}}H^{n+2},\qquad \mathscr{N}=\frac{2^{{1-n}}}{n+2},
\end{equation}
and in the lubricated region \(0 \leq r \leq r_L \) it is
\begin{subequations}
\begin{eqnarray}
\frac{\partial h}{\partial t}+\frac{1}{r}\frac{\partial
  (rq_\l)}{\partial r}&=&0,\label{eq. ndim Rey h}\\
\frac{\partial (H-h)}{\partial t}+\frac{1}{r}\frac{\partial (rq)}{\partial r}&=&0,\label{eq.q1}
\end{eqnarray}
\end{subequations}
where
\begin{subequations} \label{eq.flux}
  \begin{eqnarray} \label{eq.q}
    q&=&- \mathscr{N}
    \frac{\partial H}{\partial r} \left| \frac{\partial H}{\partial
         r}\right|^{n-1}(H-h)^{2+n}\nonumber\\
    &&- \mathscr{M} h(H-h) \left[\frac{\partial H}{\partial
        r}(H-h)+\left(\frac{\partial H}{\partial
          r}+\mathscr{D}\frac{\partial h}{\partial
          r}\right)\frac{h}{2}\right],\\
  q_\l&=&-\M \frac{h^2}{2}\left[\frac{\partial H}{\partial
      r}(H-h)+\frac{2h}{3}\left(\frac{\partial H}{\partial
        r}+\mathscr{D}\frac{\partial h}{\partial r}\right)\right], \label{eq.ql} 
\end{eqnarray}  
\end{subequations}
and where the boundary conditions are
\begin{subequations}
  \label{eq.ndima}
  \begin{eqnarray}
    q=0, \quad H=0 \quad &\textrm{at}& \quad r=r_N,\\
    q_\l=0, \quad h=0, \quad q^+=q^-, \quad H^+=H^- \quad &\textrm{at}& \quad r=r_L,
  \end{eqnarray}
  \begin{equation} \label{eq.ndime} \lim_{r\rightarrow 0} (2\pi
    rq_\l)=\alpha \mathscr{Q}\left(t-\hat{t}_L\right)^{\alpha -1}
    \quad \textrm{and} \quad \lim_{r\rightarrow 0} (2\pi rq)=\alpha
    t^{\alpha -1},
  \end{equation}
\end{subequations}
where $\hat{t}_L=t_L/T$. The resulted dimensionless quantities
\begin{equation}
   \label{eq.DQM}
   \mathscr{Q}\equiv\frac{Q_\l}{Q},\quad
   \mathscr{D}\equiv\frac{\rho _\l-\rho}{\rho},\quad
   n,\quad
   \mathscr{M}\equiv\frac{\mu}{\mu_\l}=
   \frac{\rho g}{\mu_\l}\left(\frac{k}{\rho g}\right)^{\ts{\frac{8n}{5n+3}}}
    \left(\frac{T^{5-\alpha}}{Q}\right) ^{\ts{\frac{n-1}{5n+3}}},
\end{equation}
represent respectively the discharge flux ratio, the relative density
difference of the lubricating and power-law fluids, the power-law
fluid exponent, and the dynamic viscosity ratio.

The viscosity ratio \ref{eq.DQM}(iv) implies that the timescale $T$ in
\eqref{eq:scales} is scaled out of the equations in two special cases
\(n=1\) and \(\alpha=5\). This implies that asymptotically in time
($t/t_L\gg 1$) our PDE set admits a global similarity solution of the
first kind in these two special cases, whereas for any other value of
$n$ or $\alpha$, including the constant flux case $(\alpha=1)$, such
global similarity solutions do not exist. Nevertheless, as we show in
the section that follows (\S \ref{sec:similarity}), we find several
additional asymptotic limits in which part of the flow evolves in a
self-similar manner.

In the general case $n\ne 1$ and $\alpha\ne 5$ there are enough
relations to determine the time scale \(T\) \eqref{eq:scales}
independently of the height and radial scales.  This is primarily
because the flux in the top fluid layer \eqref{eq.q} consists of two
contributions that are not necessarily of the same scale. Requiring
that the scales of those two contributions balance
$\mathscr{M}(T)=\mathscr{N}$ leads to an independent constraint for
the time scale $T$, which upon substitution in \eqref{eq:scales} leads
to a set of independent scales
\begin{subequations}
  \label{eq:TRHnewScales}
  \begin{eqnarray}
    T^{(1-n)(\alpha-5)}&=&Q^{n-1}\left(\mathscr{N}\frac{\mu_\l}{\rho g}\right)^{5n+3} \left(\frac{\rho g}{k}\right) ^{8n},\\
    R^{(1-n)(\alpha-5)}&=&Q^{2(n-1)}\left(\mathscr{N}\frac{\mu_\l}{\rho
                           g}\right)^{\a+1+2\a n} \left(\frac{\rho g}{k}\right)^{n(3\a+1)},\\
    H^{(1-n)(\alpha-5)}&=&Q^{n-1}\left(\mathscr{N}\frac{\mu_\l}{\rho
                           g}\right)^{\a-2+\a n} \left(\frac{\rho g}{k}\right) ^{2n(\a-1)}.
  \end{eqnarray}
\end{subequations}
Therefore, in the more general case $(n\ne 1$ and $\alpha\ne 5$) there
is no global similarity solution of the 1st kind, and the above scales
may represent the time, radius and thickness.

\section{Numerical Solution}\label{sec:numerical}

\begin{figure}[h]
\centering
\includegraphics[scale=0.55]{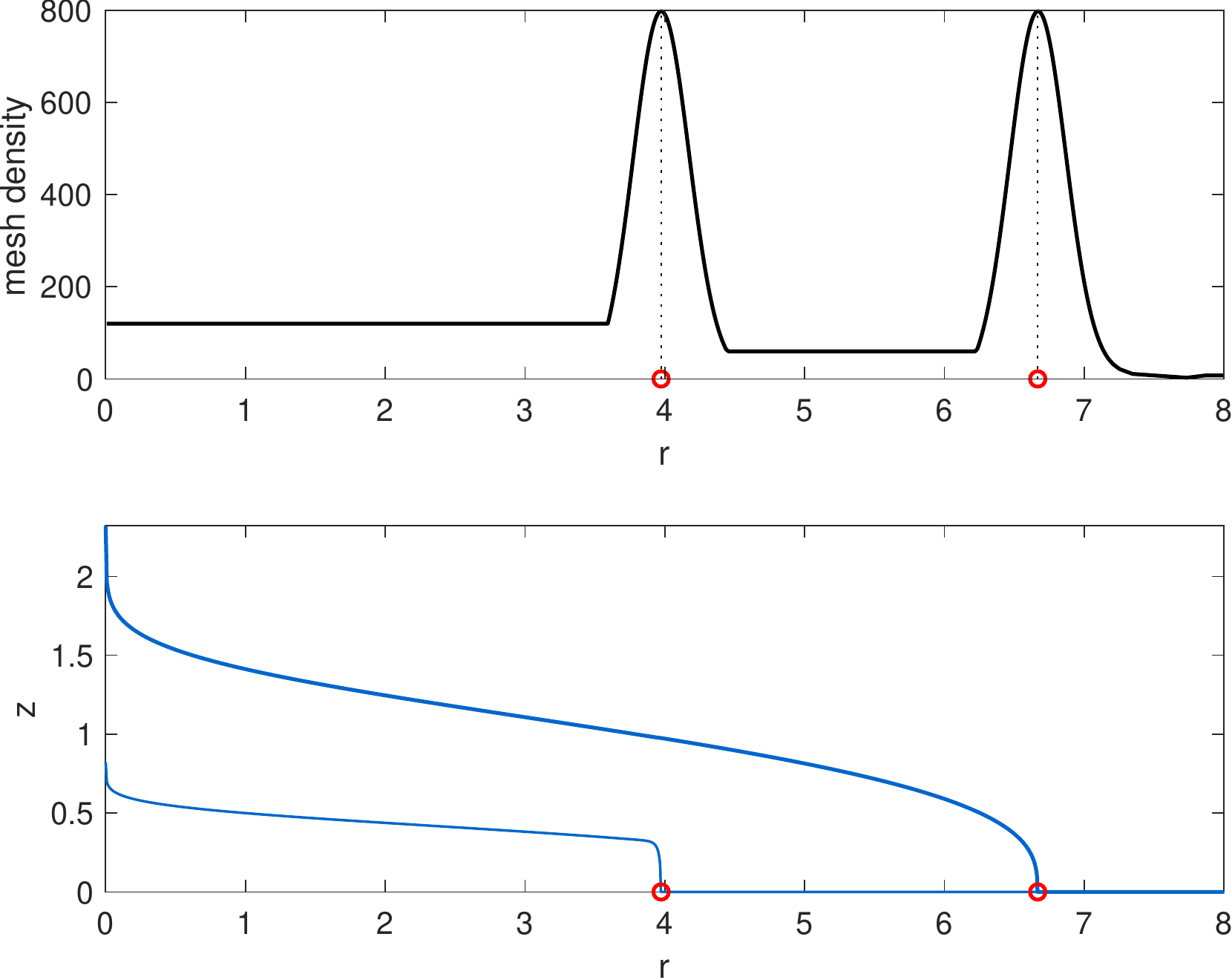}
%
\caption{(a) Mesh density and (b) a solution at time $t/t_L=100$,
  showing the lubricated fluid (thick) and the lubricating fluid
  (thin) for \(n=1\), \(\M=1\), \(\Q=0.2\), and \(\D=0.1\). Markers
  (red) indicate the front positions.
  \label{adaptmesh}}
\end{figure}

We solve the dimensionless model (eqs.~\ref{eq.ndimb} -- \ref{eq.DQM})
numerically explicitly using the Matlab PDEPE solver, with an
open-ended, non-uniform, and time-dependent adaptive spatial mesh.  We
find that lower spatial resolution can be used in most of the domain
if the mesh is logarithmically spaced, and is denser around the fluid
fronts, where the fluid heights drop sharply and the slopes
\(\p{ H}/\p{ r}\) and \(\p{ h}/\p{ r}\) become singular. Therefore, we
use a non-uniform spatial mesh that consists of two Gaussian
distributions centered at the fronts \(r_N\) and \(r_L\), and uniform
distributions between the origin and \(r_L\), the two fronts, and
beyond $r_N$ (Figure \ref{adaptmesh}). The instantaneous front
positions \(r_N(t)\) and \(r_L(t)\) are determined where
\(H(r,t) \leq 10^{-6}\) and \(h(r,t) \leq 10^{-6}\)
respectively. Since the fronts evolve, we enlarge the flow domain
progressively and re-mesh the computational grid points every several
time steps in order to keep the higher mesh density centered at the
front positions. For the re-mesh we predict the front positions in the
next time step using a discretization of the front evolution equations
\begin{equation}
  \dot{r}_{N}(t)=\lim _{r\rightarrow r_{N}} \frac{q}{H} \quad \textrm{and} \quad \Dot{r}_{L}(t)=\lim _{r\rightarrow r_{L}} \frac{q_{\l}}{h}.
\end{equation}
The size of the time step needed in order to solve the PDE is
determined by the solver, and we scale $T$ with $t_L$, so that
$\hat{t}_L=1$.

We validated our numerical solver using several known asymptotic
solutions.  In the limit \(\mathscr{Q}=0\) (no lubricating fluid) our
model converges to a GC that propagates under no-slip
condition along the substrate, which has a similarity solution
$r_N(t) \propto t^{[\a(2n+1)+1]/{(5n+3)}}$
\citep{Huppert:1982-JFM-Propagation,Sayag-Worster:2013-Axisymmetric}. We
find that our numerical solutions for the fluid heights and for the
leading front are consistent with those theoretical predictions
(Appendix \ref{sec:numerical scheme}).
Particularly, discrepancy between the predicted and computed exponents
is less than \(5\cdot 10^{-3}\) and can be minimized further with a
higher spatial resolution.
In the limit $n=1$ and $\a=1$ our model converges to a purely
Newtonian lubricated GC released at constant flux, which
has a similarity solution $r_L,r_N\propto t^{1/2}$
\citep{KowalWorster:2015-JFM-Lubricated}. We find that our numerical
solutions are consistent with both the front and thickness predictions
for a wide range of $\M$ and $\Q$ values (Appendix \ref{sec:numerical scheme}).

\section{ Similarity solutions}
\label{sec:similarity}

The model described by eqs. \ref{eq.ndimb}-\ref{eq.DQM} admits
similarity solutions in several special cases. Specifically, global
similarity solutions exist when the top fluid is also Newtonian
($n=1$), or when the mass-discharge exponent is $\alpha=5$. In
addition, part of the flow evolves in a self-similar manner in the
asymptotic limits of the viscosity ratio $\M$. Finally, for any
parameter combination the solutions at the vicinity of each front
evolve in a self-similar manner.

\subsection{The similarity solution for Newtonian fluids \(n=1\)}
\label{sec:neq1}

When the top fluid is Newtonian ($n=1$) the viscosity ratio
$\mathscr{M}$ is independent of $T$, and the resulted
purely Newtonian coupled flow admits a global similarity
solution. This case, restricted to a constant flux (\(\alpha=1\)), has
been thoroughly explored \citep{KowalWorster:2015-JFM-Lubricated}. For
the general case of \(\alpha \geq 0\), the PDE set can be reduced to
an ODE set with a similarity variable
\begin{equation} 
    \eta=\frac{r}{t^{\frac{3\alpha +1}{8}}}\left(\frac{\rho g}{k} Q^{3}\right)^{-\frac{1}{8}},
\end{equation}
and a solution of the form
\begin{subequations}
\label{eq:similarity n=1}
\begin{equation}
    \left(H,h\right)=t^{\frac{\alpha-1}{4}}\left(Q \frac{k}{\rho g}\right)^{\frac{1}{4}}\left(F,f\right),
\end{equation}
where \(F\) and \(f\) are dimensionless functions of
\(\eta\). Therefore, the fronts of both the lubricated and lubricating
fluids evolve like
\begin{equation}
  \label{rLN n=1}
    \left(r_N,r_L\right)=\left(\eta_N,\eta_L\right)t^{\frac{3\alpha+1}{8}} \left( Q^{3} \frac{\rho g}{k} \right)^{\frac{1}{8}},
  \end{equation}
\end{subequations}
where \(\eta_N\) and \(\eta_L\) are numerical coefficients of order
\(1\). These solutions have identical structure as the classical
solution of Newtonian GCs
\citep{Huppert:1982-JFM-Propagation}.  Upon substitution, the Reynolds
equations (\ref{eq.reyNL} and \ref{eq:reyL}) become, for the
non-lubricated region ($\eta_L\le\eta\le\eta_N$)
\begin{equation}
  \left(\frac{\alpha-1}{4}\right)F-\left(\frac{3\alpha+1}{8}\right)\eta F'=\frac{1}{3\eta}\left(\eta F'F^3\right)',
\end{equation}
where prime denotes a derivative with respect to \(\eta\), and for the
lubricated region ($0\le\eta\le\eta_L$)
\begin{subequations}
\begin{eqnarray}
  \left(\frac{\alpha-1}{4}\right)\left(F-f\right)-\left(\frac{3\alpha+1}{8}\right)\eta\left(F'-f'\right) +\frac{1}{\eta}\left(\eta q\right)'&=&0,\\
  \left(\frac{\alpha-1}{4}\right)f-\left(\frac{3\alpha+1}{8}\right)\eta f' +\frac{1}{\eta}\left(\eta q_\l\right)'&=&0,
\end{eqnarray}
\end{subequations}
where
\begin{subequations}
\begin{eqnarray}
  q &=&-\frac{1}{3}F'(F-f)^3-\mathscr{M}f(F-f)\left[F'(F-f)+\frac{1}{2}f\left(\mathscr{D}f'+F'\right)\right],\\
  q_\l &=&-\mathscr{M}{f^2}\left[\frac{1}{2}F'(F-f)+\frac{1}{3}f\left(\mathscr{D}f'+F'\right)\right].
\end{eqnarray}
\end{subequations}
The corresponding boundary conditions (\ref{eq.ndima}) become
\begin{subequations}
\begin{eqnarray}
F=0, \quad q=0 \quad &\textrm{at}& \quad \eta=\eta_N\\ 
f=0, \quad q_\l=0, \quad F^+=F^-, \quad q^+=q^- \quad &\textrm{at}& \quad\eta=\eta_L,
\end{eqnarray}
\begin{equation}
  \lim _{\eta \rightarrow 0} 2 \pi \eta q = \alpha, \quad \lim _{\eta \rightarrow 0} 2 \pi \eta q_\l = \alpha \mathscr{Q},
\end{equation}
\end{subequations}
where the last condition implies convergence to a similarity solution
long time after the initiation of the fluid discharge, \(t \gg t_L\).

For $\alpha=1$ the model converges to that of
\citet{KowalWorster:2015-JFM-Lubricated}, and the similarity solutions
we predict for general $\a$ are consistent with our full numerical
solution (Figures \ref{figure04a},~\ref{figure_frontSimilarity}).

\subsection{The similarity solution for mass-discharge exponent \(\alpha = 5\)} \label{sec:a5}

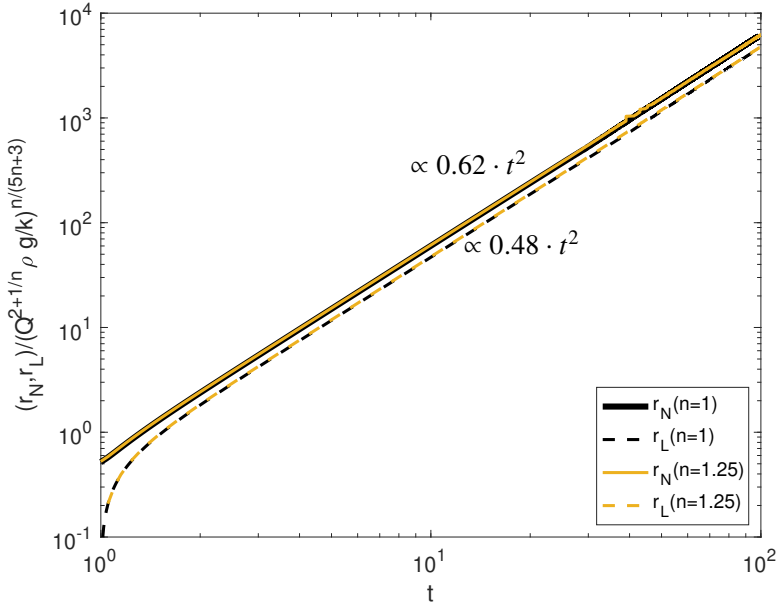
\begin{figure}
\centering
\includegraphics[scale=0.57]{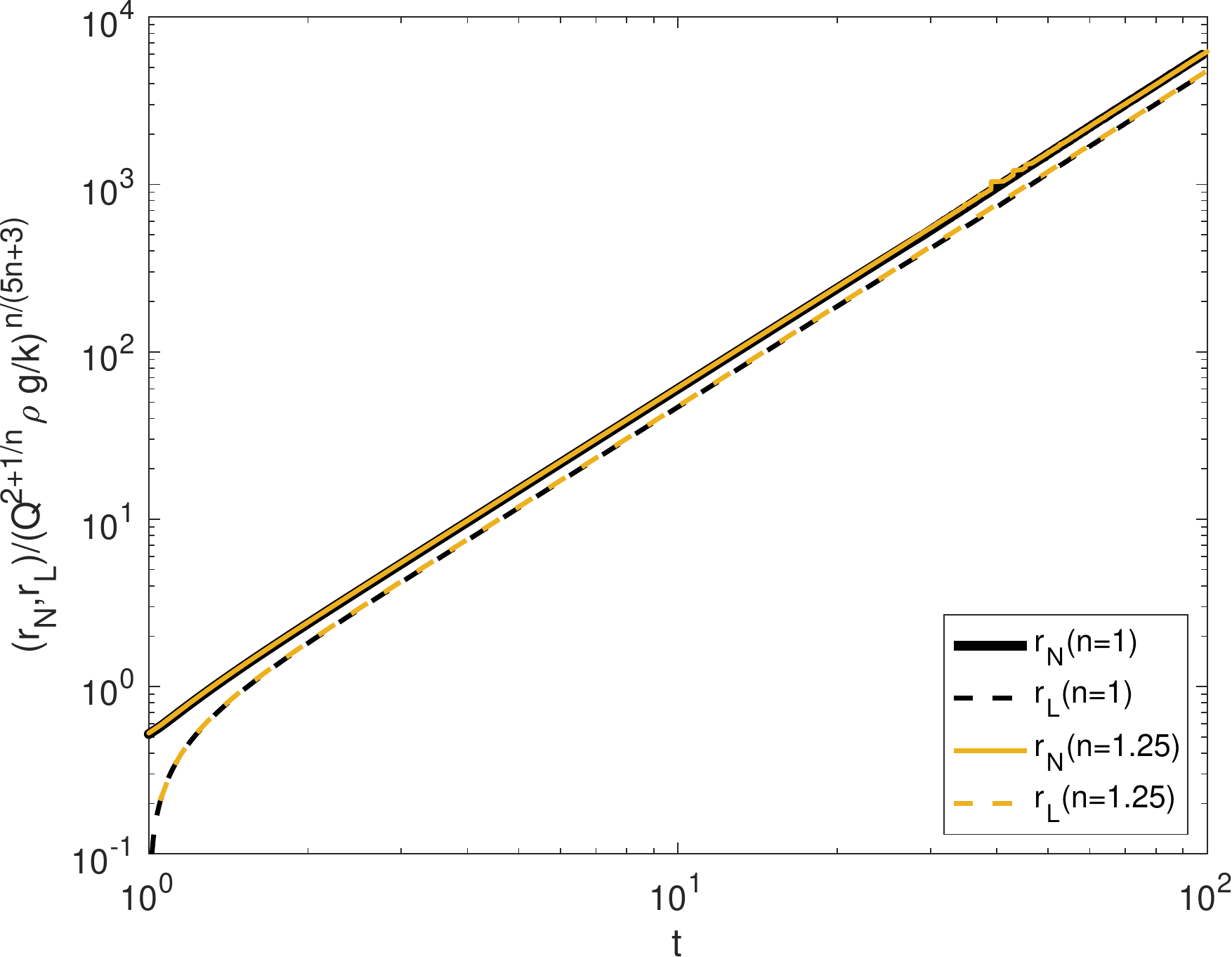}
\begin{picture}(0,0)(20,20)
   \put(-120,180){$\propto 0.62 \cdot t^{2}$}
   \put(-100,150){$\propto 0.48 \cdot t^{2}$}
\end{picture}
\caption{The front propagation verses time for \(\alpha=5\), \(n=1.25\)
  and \(1\), \(\mathscr{Q}=0.2\), \(\mathscr{D}=0.1\) and
  \(\mathscr{M}=100\) with curve regression line values indicated on
  figure, showing that a global similarity solution exists for
  \(\alpha=5\), which is consistent with theoretical prediction of the
  form $(\eta_N,\eta_L) t^2$ (eq. \ref{eq:similarity a=5}b).
  \label{figure04a}}
\end{figure}
   
When the top fluid is non-Newtonian \(n\neq 1\) the viscosity ratio
\(\mathscr{M}\) becomes independent of the time scale \(T\) for a
discharge exponent \(\alpha=5\), and the resulted flow also admits a
similarity solution with a similarity variable
\begin{equation}
\eta =\frac{r}{t^{2}} \left[ Q^{2n+1} \left(\frac{\rho g}{k}\right)^n \right]^{{-1}/{(5n+3)}}
\end{equation}
and a solution of the form
\begin{subequations}
 \label{eq:similarity a=5}
\begin{equation}
(H,h)= t\left[Q^{1+n} \left(\frac{k}{\rho g}\right)^{2n} \right]^{{1}/{(5n+3)}} (F,f),
\end{equation}
where \(F\) and \(f\) are dimensionless functions of
\(\eta\). Therefore, the fronts of both the lubricated and lubricating
fluids evolve like
\begin{equation} 
    \label{rLN a=5}
    \left(r_N,r_L\right)=\left(\eta_N,\eta_L\right)t^2 \left[ Q^{2n+1} \left(\frac{\rho g}{k}\right)^n \right]^{{1}/{(5n+3)}},
  \end{equation}
\end{subequations}
where \(\eta_N\) and \(\eta_L\) are numerical coefficients of order
\(1\). Upon substitution, the Reynolds equations (\ref{eq.reyNL} and
\ref{eq:reyL}) become, for the non-lubricated region
($\eta_L\le\eta\le\eta_N$)
\begin{equation}
F-2F'\eta=\mathscr{N}\frac{1}{\eta} \left(\eta F' \left|F'\right|^{{n-1}}F^{{n+2}} \right)',
\end{equation}
and for the lubricated region ($0\le\eta\le\eta_L$)
\begin{subequations}
\begin{eqnarray}
\left(F-f\right)-2\left(F'-f'\right)\eta+\frac{1}{\eta}\left(\eta q \right)'=0,\\
f-2f'\eta+\frac{1}{\eta}\left(\eta q_\l \right)'=0,
\end{eqnarray}
\end{subequations}
where
\begin{subequations}
\begin{eqnarray}
q&=& -\mathscr{N}F' \left|F'\right|^{n-1}\left(F-f\right)^{n+2}-\mathscr{M} f (F-f) \left[ F'(F-f)+\frac{1}{2}f (\mathscr{D}f'+F') \right],\qquad\\
q_\l&=&-\mathscr{M}{f^2 }\left[ \frac{1}{2}F'(F-f)+\frac{1}{3}f (F'+\mathscr{D}f') \right].
\end{eqnarray}
\end{subequations}
The corresponding boundary conditions (\ref{eq.ndima}) become
\begin{subequations}
\begin{eqnarray}
F=0, \quad q=0 \quad &\textrm{at}& \quad \eta=\eta_N\\
f=0, \quad q_\l=0, \quad F^+=F^-, \quad q^+=q^- \quad &\textrm{at}& \quad\eta=\eta_L,
\end{eqnarray}
\begin{equation}
\lim _{\eta \rightarrow 0} 2 \pi \eta q = 5, \quad \lim _{\eta \rightarrow 0} 2 \pi \eta q_\l = 5 \mathscr{Q}.
\end{equation}
\end{subequations}



Figure \ref{figure04a} shows numerical solution of the full PDE set
for \(\alpha=5\) for varying power-law exponent values. We find that
the numerical solutions are consistent with the predicted similarity
solution \ref{eq:similarity a=5}. This consistency further validates
the numerical simulation for the combination of lubricated gravity
currents with lubricated power-law fluid.

\subsection{Similarity solutions for the asymplotic limits in
  $\mathscr{M}$:\\ The upper-fluid solid and liquid limits}
\label{sec:asymptotic M}
In the general case of \(\alpha \neq 5\) and \(n \neq 1\) our model
does not admit global similarity solutions of the first kind. However,
a similarity solution arises in part of the flow domain in each of the
asymptotic limits of the viscosity ratio, in which the upper fluid
layer is either relatively more viscous (\(\mathscr{M} \gg 1\), the
``solid'' limit) or relatively less viscous (\(\mathscr{M} \ll 1\),
the ``liquid'' limit) compared with the lower fluid layer.

\subsubsection{The top-layer solid limit, $\mathscr{M} \gg 1$
} \label{sec.icesheet}
The limit \(\mathscr{M} \gg 1\) represents the case where the top
fluid is much more viscous than the lubricating fluid. One motivation
to study this limit is the geophysical setting of ice sheets creeping
over less-viscous lubricated beds. In this case, the leading-order
scalling of the power-law fluid local flux in the lubricated region
(\ref{eq.q}) is
\begin{equation} \label{eq054} q \approx - \mathscr{M}h(H-h)\left[\pfrac{ H}{r}(H-h)+\left(\pfrac{ H}{r}+\mathscr{D}\pfrac{ h}{r}\right)\frac{h}{2}\right],
\end{equation}
which is identical to the scaling of the Newtonian lubricating fluid
flux (\ref{eq.ql}). In such a case, the three scales \(H\), \(R\) and
\(T\) in the lubricated region cannot be determined independently
despite the fact that the upper fluid is a power-law fluid. Therefore,
both fluids in the lubricated region have a similarity solution in
which the dimensionless front in the lubricated region evolves with an
$n$-independent exponent
\begin{equation} \label{eq.Mgg1lim}
r_L=\xi_L t^{\frac{3\alpha+1}{8}},
\end{equation}
where the intercept \(\xi_L\) may depend on the system dimensionless
numbers. The exponent in (\ref{eq.Mgg1lim}) is consistent with the
predictions made for the special cases discussed above. Specifically,
for \(n=1\) it is \((3\alpha+1)/8\) as we predict in equation
\eqref{rLN n=1}, and for \(\alpha=5\) it is \(2\) independently of $n$
as we predict in equation \eqref{rLN a=5}. This similarity solution
does not reveal the evolution of the fluid front $r_N$ in the
non-lubricated region, which depends on the fluid exponent $n$.

\subsubsection{The top-layer ``liquid'' limit, $\mathscr{M} \ll
  1$} \label{sec.invicid}

The second asymptotic limit, \(\mathscr{M} \ll 1\), in which the top
fluid is much less viscous than the lubricating fluid, also admits a
similarity solution, but of a different sub set of the model. In this
case, the leading-order scaling of the power-law fluid local flux in
the lubricated region (\ref{eq.q}) is
\begin{equation}
  \label{eq:qMll1}
  q \approx - \mathscr{N} \frac{\partial
    H}{\partial r} \left| \frac{\partial H}{\partial
      r}\right|^{{n-1}}(H-h)^{{2+n}},
\end{equation}
which is identical to the scaling of the power-law fluid in the
non-lubricated region \eqref{eq.q_noslip}. Therefore, the evolution of
the power-law fluid in the entire domain converges when $h\ll H$ to
the similarity solution of a non-lubricated power-law GC,
in which case the upper fluid front evolves with an $n$-dependent
exponent
\begin{equation} \label{simMll}
r_N=\xi_N\mathscr{N}^{1/(5n+3)} t^{\ts{\frac{\a(2n+1)+1}{5n+3}}}
\end{equation}
\citep{Sayag-Worster:2013-Axisymmetric}, where the intercept \(\xi_N\)
may depend on the system dimensionless numbers.  This similarity
solution does not hold for the lubricating fluid layer since the local
flux of the Newtonian lubricating fluid sets a different
scaling. Therefore, it does not predict the evolution of the
lubricating-fluid front $r_L$. For  $H-h>1$ the coupling between
the upper fluid layer and the lower one grows stronger the more
shear-thinning the fluid is ($n\rightarrow \infty$), as the exponent in
\eqref{eq:qMll1} grows like $n$. Therefore, \eqref{simMll} is
expected to be less accurate the more shear-thinning the upper fluid
is when $H-h>1$, and more accurate when $H-h<1$.

\subsection{Similarity solutions at the vicinity of the fluid fronts}
\label{sec:similarity fronts}

\begin{figure}
  \centering 
  \includegraphics[scale=0.45]{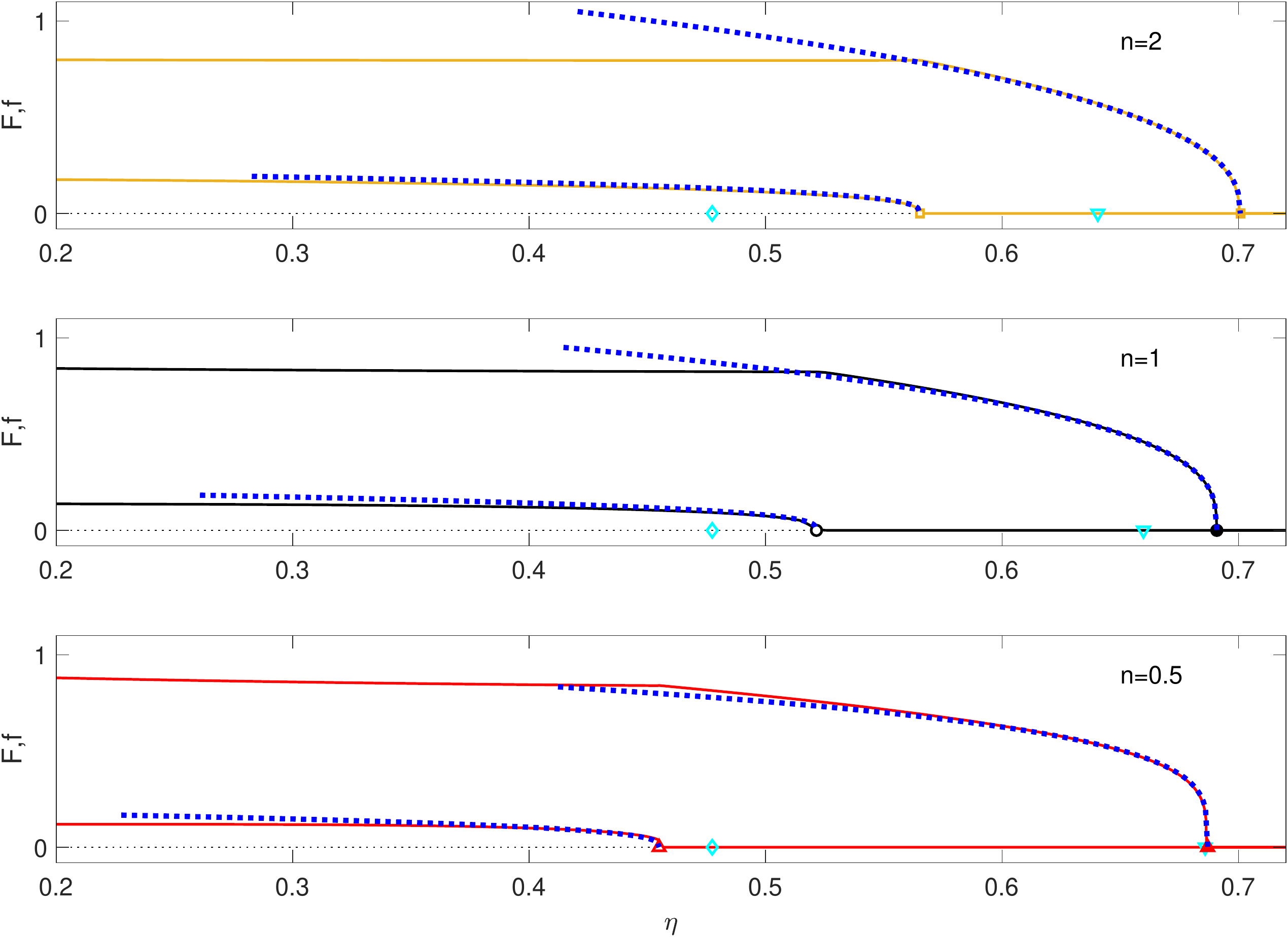}
\begin{picture}(0,0)(20,20)
\end{picture}
\caption{\label{figure_frontSimilarity} The similarity solutions at
  the vicinity of the fronts $F(\eta),f(\eta)$ (eqs.~\ref{eq:front H},
  \ref{eq:front h}) based on the non-lubricated theory
  (\textcolor{blue}{\tiny$\bullet\bullet\bullet$}, blue), compared
  with the full numerical solution (---) for $\M=100,\Q=0.1,\D=1,\a=1$
  and $n=0.5,1,2$. The similarity solutions were plotted with respect
  to the lubricated-theory fronts (${\square}, \circ,{\triangle}$
  markers at corresponding colors to the $F,f$ curves). For comparison
  the fronts predicted by the non-lubricated theory are marked
  (\textcolor{cyan}{$\bigtriangledown$} (cyan) marks $\eta_N$ and
  \textcolor{cyan}{$\Diamond$} (cyan) marks $\eta_L$). }
\end{figure}

The thicknesses of the two fluid layers at the vicinity of the fronts
have also got self-similar forms, even at the absence of a global
similarity solution. These self-similar forms arise because the
dynamics at the vicinity of both fronts is dominated by the same
dynamics that governs non-lubricated GCs, which are known
to have self-similar solutions for any $\a$ and $n$
\citep{Sayag-Worster:2013-Axisymmetric,Huppert:1982-JFM-Propagation}.

The dominance of a non-lubricated-GC dynamics is naturally expected at
the vicinity of the front $r_N$, which is the edge of the
non-lubricated region in the flow. In this case we expect consistency
of the fluid thickness with \citet{Sayag-Worster:2013-Axisymmetric},
which implies a similarity solution of the form
\begin{subequations}
  \begin{eqnarray}
  \label{eq:front rN solution}
  r&=&\eta t^{\ts{\frac{\a(2n+1)+1}{5n+3}}},\\
  H&=&t^{\ts{\frac{\a(n+1)-2}{5n+3}}}F\left(\frac{\eta}{\eta_N}\right)
  \end{eqnarray}
\end{subequations}
for the model described by equations (\ref{eq.ndimb},
\ref{eq.q_noslip}), where $\eta$ is the similarity variable and
$\eta_N\equiv\eta(r=r_N)$.  Therefore, the asymptotic solution for the
function $F$ near the front $\eta_N$ in the similarity space is
\citep{Sayag-Worster:2013-Axisymmetric}
\begin{equation}
  \label{eq:front H}
  F=\left[N\eta_N^{n+1}\left(1-\frac{\eta}{\eta_N}\right)^n\right]^{{1}/{(2n+1)}},\qquad
  N=\left(\frac{1}{\N}\frac{\a(2n+1)+1}{5n+3}\right)^{1/n}\frac{2n+1}{n}.   
\end{equation}
Near the front $r_L$ the dominance of non-lubricated-GC dynamics is
also valid, but requires more careful supporting
argument. Specifically, we assert that the free surface slope
$\p H/\p r$ is finite at $r_L$, whereas the slope of the lubricating
fluid is singular. Therefore, provided the density difference is non
zero, the leading-order flux $q_\l$ \eqref{eq.ql} is
\begin{equation}
  q_\l=-\frac{\M\D}{3}h^3\pfrac{h}{r}.
\end{equation}
Consequently, the model \eqref{eq. ndim Rey h} for the lubricating
fluid near $r_L$ is decoupled from $H$, and describes a modified
non-lubricated GC of a Newtonian fluid
\citep{Huppert:1982-JFM-Propagation}. In this case the thickness $h$
has a similarity solution of the form
\begin{subequations}
  \begin{eqnarray}
  \label{eq:front rL solution}
  r&=&\eta (t-t_L)^{\ts{\frac{3\a+1}{8}}},\\
  h&=&(t-t_L)^{\ts{\frac{\a-1}{4}}}f\left(\frac{\eta}{\eta_L}\right),
  \end{eqnarray}
\end{subequations}
where $\eta_L\equiv\eta(r=r_L)$.  Therefore, the asymptotic solution
for the function $f$ near the front $\eta_L$ in the similarity space
is \citep{Huppert:1982-JFM-Propagation}
\begin{equation}
  \label{eq:front h}
  f=\left[ \frac{9(3\a+1)}{8}\frac{\eta _L^2}{\M\D} \left( 1 - \frac{\eta}{\eta_L} \right) \right]^{1/3}.
\end{equation}
Plugging the similarity solutions for $F$ and $f$ into the dimensionless form of the
global mass conservation equations \ref{eq:vol} in the late time limit
$t/t_L\gg 1$
\begin{subequations}
\begin{equation}
2 \pi \int_{0}^{\eta_N} (F-f)\eta d\eta=1,
\end{equation}
\begin{equation}
2 \pi \int_{0}^{\eta_L} f \eta d\eta=\Q,
\end{equation}
\end{subequations}
we obtain the closed-form solutions for the fronts intercept
\begin{subequations}
  \label{eq:NL etaNL}
\begin{eqnarray}
\eta_N&=&\left[\frac{2\pi}{1+\Q}\frac{(2n+1)^2}{(5n+2)(3n+1)} N^{\ts{\frac{n}{2n+1}}} \right]^{\ts{-\frac{2n+1}{5n+3}}},\\
\eta_L&=&\left[\frac{2\pi}{\Q}\frac{9}{28} \left(\frac{9(3\a+1)}{8\M\D} \right)^{\frac{1}{3}} \right]^{-\frac{3}{8}}.
\end{eqnarray}
\end{subequations}
The asymptotic solutions \ref{eq:front H} and \ref{eq:front h} that we
obtain based on the non-lubricated theory appear to provide a precise
prediction to the fluid thicknesses of the lubricated GC at the
vicinity of the fronts when using the numerically calculated values
for $\eta_L,\eta_N$ of the lubricated GC (Figure
\ref{figure_frontSimilarity}). The values for $\eta_L$ and $\eta_N$
\eqref{eq:NL etaNL} based on the non-lubricated GC theory provide a
rough estimate for those of the lubricated theory.  Discrepancies
between the two seem to decline the more shear-thickening the upper
fluid is (Figure \ref{figure_etaNL}a).

\section{Front outstripping}
\label{sec:outstripping}

In the absence of similarity solutions the fronts propagate at
different rates, implying that under certain conditions the inner
lubricating front $r_L$ can outstrip the outer front $r_N$. We
identify two outstripping mechanisms, one driven by the solution
intercepts and the other by the exponents.

\subsection{Intercept outstripping}
\label{sec:outstripping:intercept}

\begin{figure}
 \includegraphics[scale=0.4]{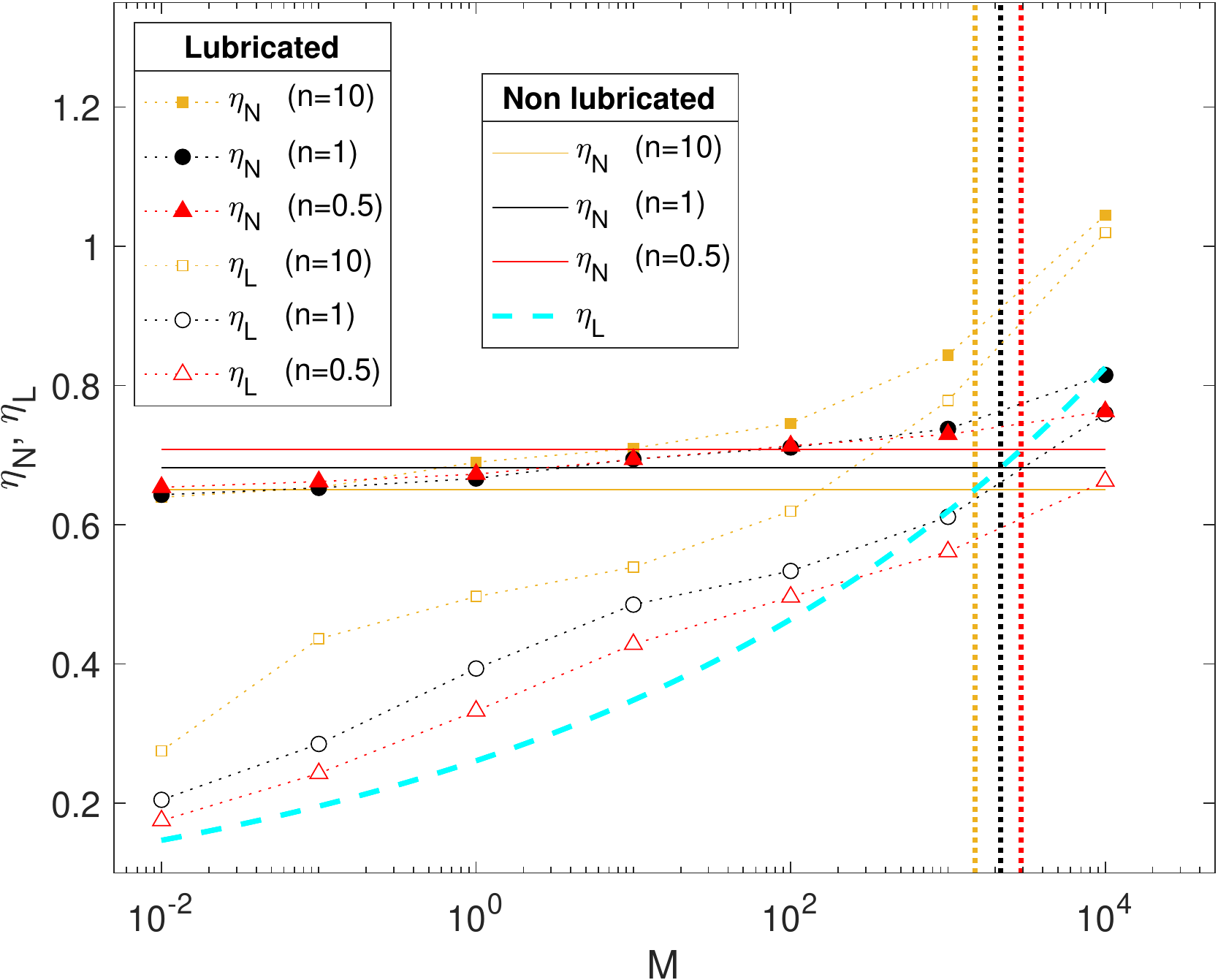}
 \includegraphics[scale=0.4]{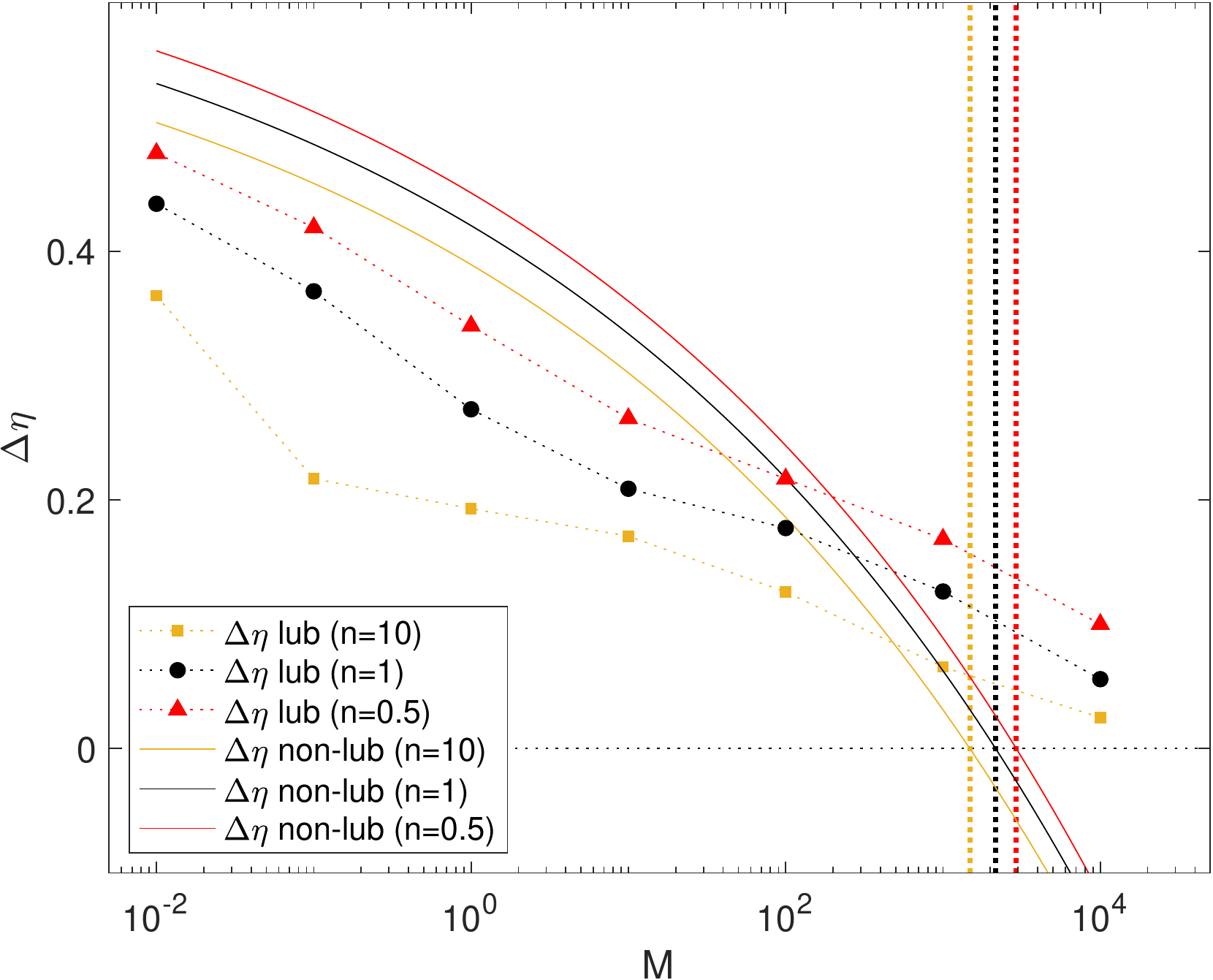}
\begin{picture}(0,0)(20,20)
   \put(20,30){(a)}
   \put(215,30){(b)}
\end{picture}
\caption{\label{figure_etaNL} (a) The coefficients $\eta_N,\eta_L$ of the
  similarity solutions at the vicinity of the fronts of the lubricated
  GC compared with the non-lubricated solutions, where
  $\Q=0.2, \D=0.1, \a=1$, $10^{-2}\le\M\le 10^4$, and $n=10,1,0.5$. The
  $\eta_N$ of the non-lubricated theory (solid lines) are independent
  of $\M$ and diminish with $n$, whereas those of the lubricated theory
  (solid markers at corresponding colors) are nearly equal in the
  liquid limit ($\M\ll 1$) and grow monotonically with $\M$. 
  The corresponding $\eta_L$ of the non-lubricated theory (- -
  -, cyan) are independent of $n$, unlike  those of the lubricated
  theory (hollow markers),  but both grows monotonically with
  $\M$.
  Vertical grid lines mark the non-lubricated threshold viscosity
  ratio $\M_c$.
  (b) The difference $\Delta\eta\equiv\eta_N-\eta_L$ of the lubricated
  and non-lubricated values in panel a. }
\end{figure}
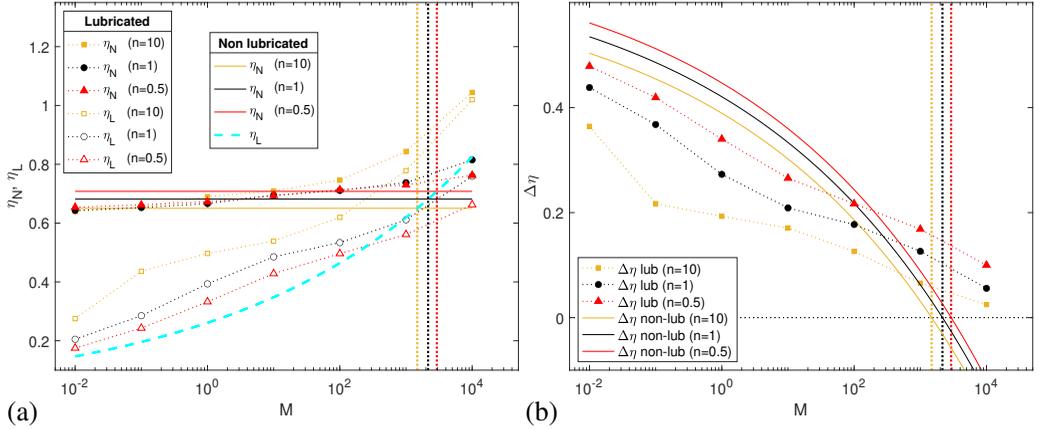

The intercept difference $\Delta\eta\equiv\eta_N-\eta_L$ predicted by
the non-lubricated theory \eqref{eq:NL etaNL} diminishes with $\M$ and
becomes negative when $\M$ grows beyond a threshold value that we
denote by $\M_c$ (Figure \ref{figure_etaNL}b). This implies that
across that threshold in the viscosity ratio \(\eta_L>\eta_N\) and the
lubricating front outstrips the upper fluid front.  The threshold
viscosity ratio $\M_c$ is found by setting $\Delta\eta=0$ and solving
for $\M$
\begin{equation}
    \label{eq:Mc def}
\M_c\equiv\left(\frac{2\pi}{\Q}\frac{9}{28}\right)^3 \left(\frac{9(3\a+1)}{8\D} \right)\left[\frac{1+\Q}{2\pi}\frac{(5n+2)(3n+1)}{(2n+1)^2} N^{\ts{-\frac{n}{2n+1}}} \right]^{\ts{\frac{8(2n+1)}{5n+3}}}.
\end{equation}
Therefore, in the purely Newtonian case the threshold viscosity ratio
is
\begin{eqnarray}
  \label{eq:Mc n=1}
  \M_c(n=1)&=&  \frac{1}{\D}\left(1+\frac{1}{\Q}\right)^3
\end{eqnarray}
and is independent of $\a$. In the case of constant flux $\a=1$ the
threshold is
\begin{eqnarray}
  \label{eq:Mc a=1}
  \M_c(\a=1)&=& \left(\frac{2\pi}{\Q}\frac{9}{28}\right)^3 \frac{9}{2\D} \left[\frac{1+\Q}{2\pi}\frac{(5n+2)(3n+1)}{(2n+1)^2} N^{\ts{-\frac{n}{2n+1}}} \right]^{\ts{\frac{8(2n+1)}{5n+3}}}.
\end{eqnarray}
This behaviour reasoned based on the non-lubricated theory is
consistent with the trend implied from the numerical solution of the
full lubricated theory (Figure \ref{figure_etaNL}b), though we do not
explicitly observe an outstripping of the outer front by the inner one
since the present simulation assumes $r_L<r_N$.

\subsection{Exponent outstripping}
\label{sec:outstripping:exponent}

\begin{figure}
  \centering
  \vspace{-1mm}
\includegraphics[scale=0.48]{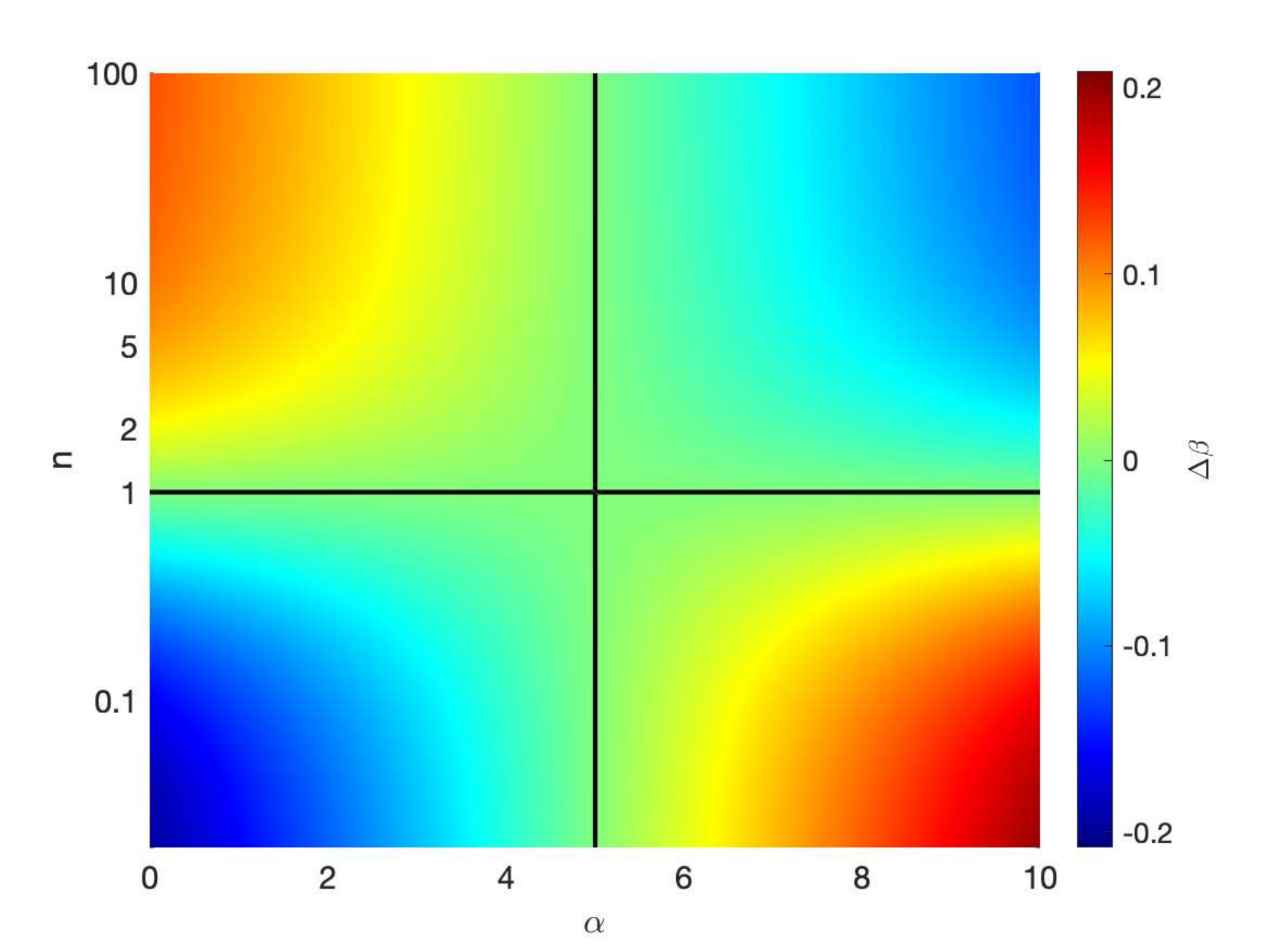}
\begin{picture}(0,0)(20,20)
  \put(-122,120){\small $n=1$ similarity solution}
  \put(-135,120){\rotatebox{90}{\small $\alpha=5$ similarity solution}}
  \put(-100,75){\small \textcolor{gray}{Outstripping}}
  \put(-200,160){\small \textcolor{gray}{Outstripping}}
\end{picture}
  \vspace{-3mm}
\caption{\label{figure_outstrip_exp} The exponent difference
  $\Delta\beta$ as a function of the fluid discharge exponent $\alpha$
  and the viscosity exponent $n$. Dark lines are the $\Delta\beta=0$
  contours, where global similarity solutions exist.  When
  $\Delta\beta>0$ (hot colors) the lubricating fluid front $r_L$
  outstrips the non-Newtonian fluid front $r_N$. }
\end{figure}
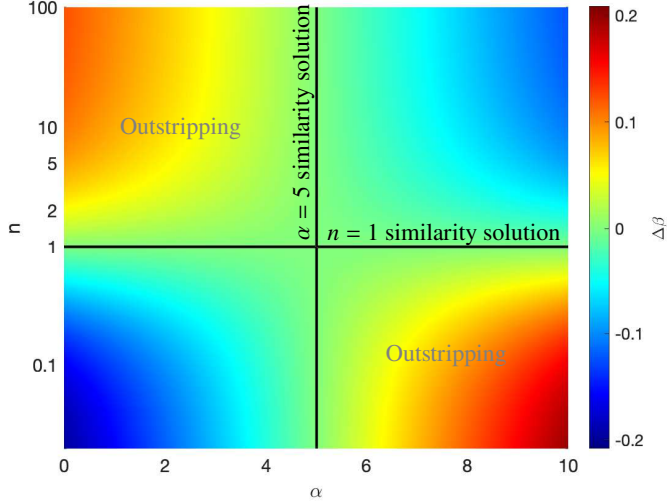

Another mechanism of outstripping can arise due to the different time
exponents of the two fronts. This difference can be evaluated from the
similarity solutions at the vicinity of each front. Specifically,
denoting the front exponents by $\beta_N\equiv[\a(1+2n)+1]/(5n+3)$
(\ref{eq:front rN solution}) and $\beta_L\equiv(3\a+1)/8$
(\ref{eq:front rL solution}) then
\begin{equation}
  \label{eq:dphi}
  \Delta\beta\equiv\beta_L-\beta_N= \frac{(5-\alpha)(n-1)}{8(5n+3)}.
\end{equation}
When the flow has a global similarity solution ($\alpha=5$ or $n=1$)
the exponents of the two fluid fronts are identical $\Delta\beta=0$,
implying that asymptotically in time the gap between the two fronts
evolves with the same exponent and the ratio of their position
$r_L/r_N$ is constant. When $n\ne 1$ and $\alpha\ne 5$ there is no
global similarity solution, implying that $\beta_L\ne\beta_N$ and that
the fronts ratio evolves proportionally to
$t^{\Delta\beta}$. Consequently, the gap between the fronts closes
down in time and front outstripping occurs when $\Delta\beta>0$, which
is when $\alpha<5 ~\&~ n >1$ or when $\alpha>5 ~\&~ n <1$ (Figure
\ref{figure_outstrip_exp}).  When $\Delta\beta<0$ intercept
outstripping can still occur at early time $(t/t_L\gtrsim 1)$ when
$\M>\M_c$.

\section{Solutions for constant-flux discharge \(\alpha=1\)}
\label{sec:const}

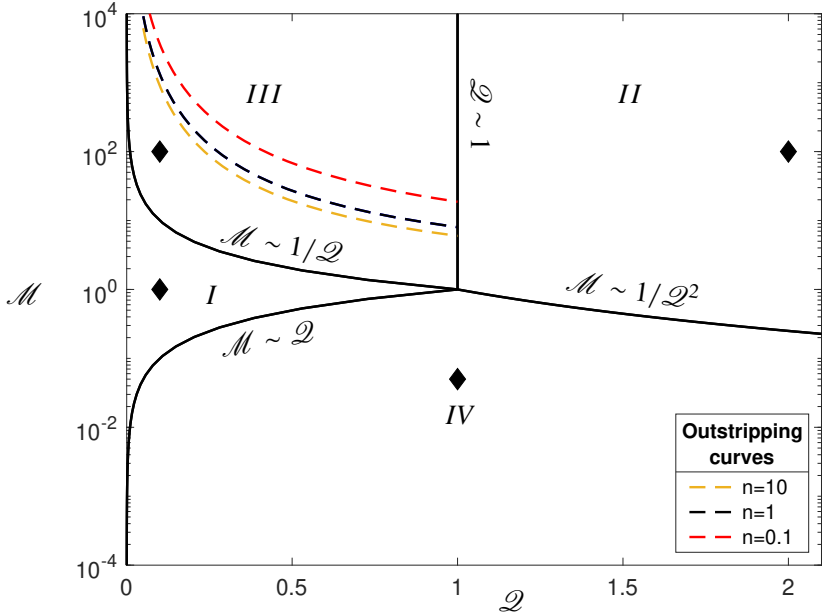
\begin{figure}
  \centering 
   \includegraphics[scale=0.6]{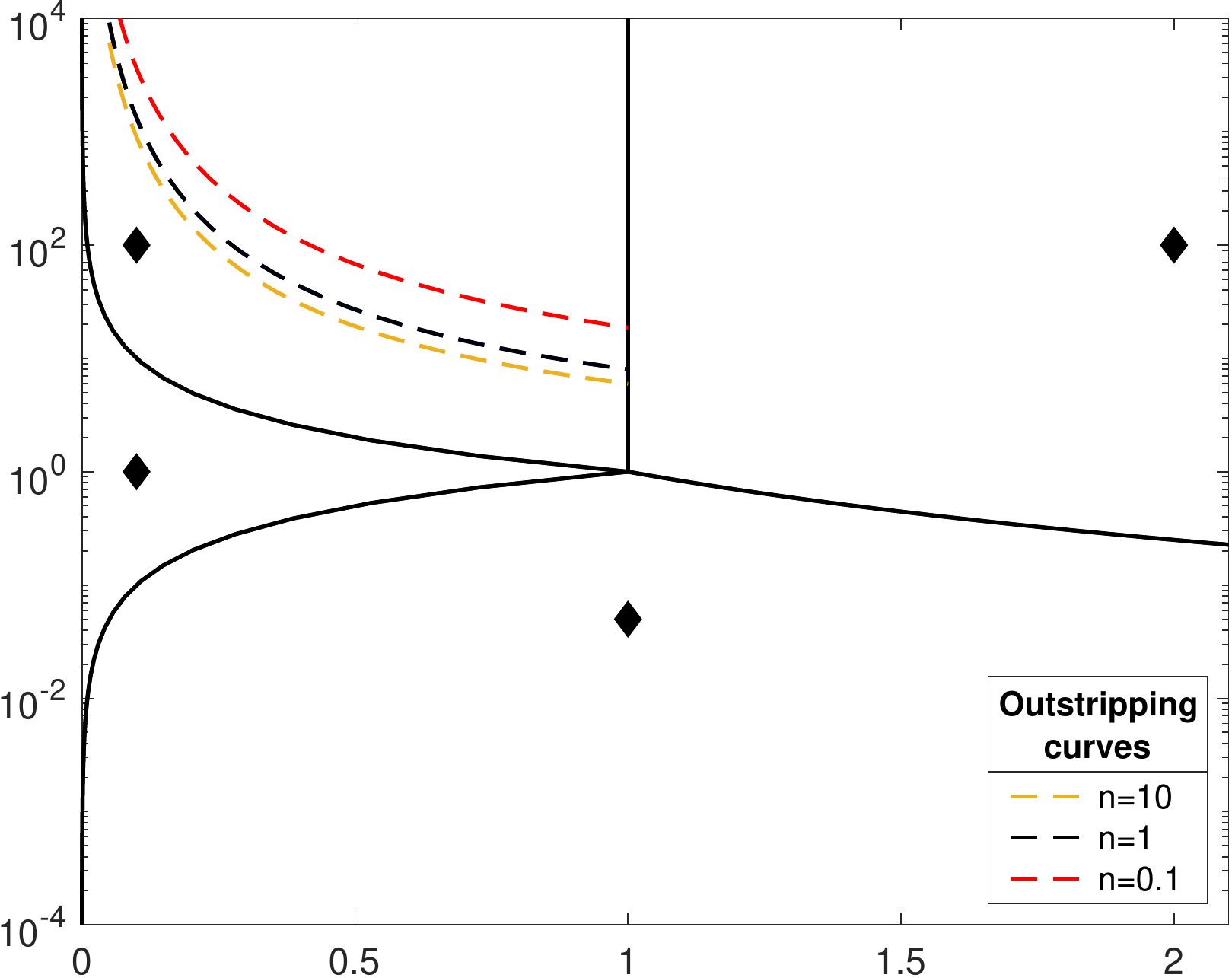}
\begin{picture}(0,0)(-10,35)
   \put(-320,145){$\M$}
   \put(-135,30){$\Q$}
   \put(-245,145){$I$}
   \put(-90,220){$II$}
   \put(-230,220){$III$}
   \put(-155,100){$IV$}
   \put(-240,125){\rotatebox{10}{$\M \sim \Q$}} 
   \put(-110,150){\rotatebox{-8}{$\M \sim 1/\Q^{2}$}}
   \put(-145,230){\rotatebox{-90}{$\Q \sim 1$}}
   \put(-240,168){\rotatebox{-10}{$\M \sim 1/\Q$}}
 \end{picture}
\caption{ The \(\M\) - \(\Q\) state map for $\alpha=1$, $n=1$ and
  $\D=1$ is divided into four characteristic flow regimes
  \textit{I-IV}.  The dashed curves marks the critical viscosity ratios
  \(\M_c(\Q,n)\) (equation \ref{eq:Mc a=1}), beyond which we expect
  the front of the lubricating fluid to outstrip the front of the top
  fluid. The characteristic solution of the \(\M-\Q\) states marked by
  diamonds are shown in Figure \ref{figure3}.
  \label{fig:M-Q state map}}
\end{figure}

The case of constant flux (\(\alpha=1\)) was explored theoretically in
several asymptotic limits, including the case of no lubrication
\citep{Huppert:1982-JFM-Propagation,Sayag-Worster:2013-Axisymmetric}
and the case where both fluids are Newtonian
\citep{KowalWorster:2015-JFM-Lubricated}.  These cases are useful to
validate our general solutions in several asymptotic limits and to
elucidate the impact of lubrication when the upper fluid is
non-Newtonian.

\begin{figure}
\begin{subfigure}{0.5\textwidth}
\centering
\includegraphics[scale=0.37]{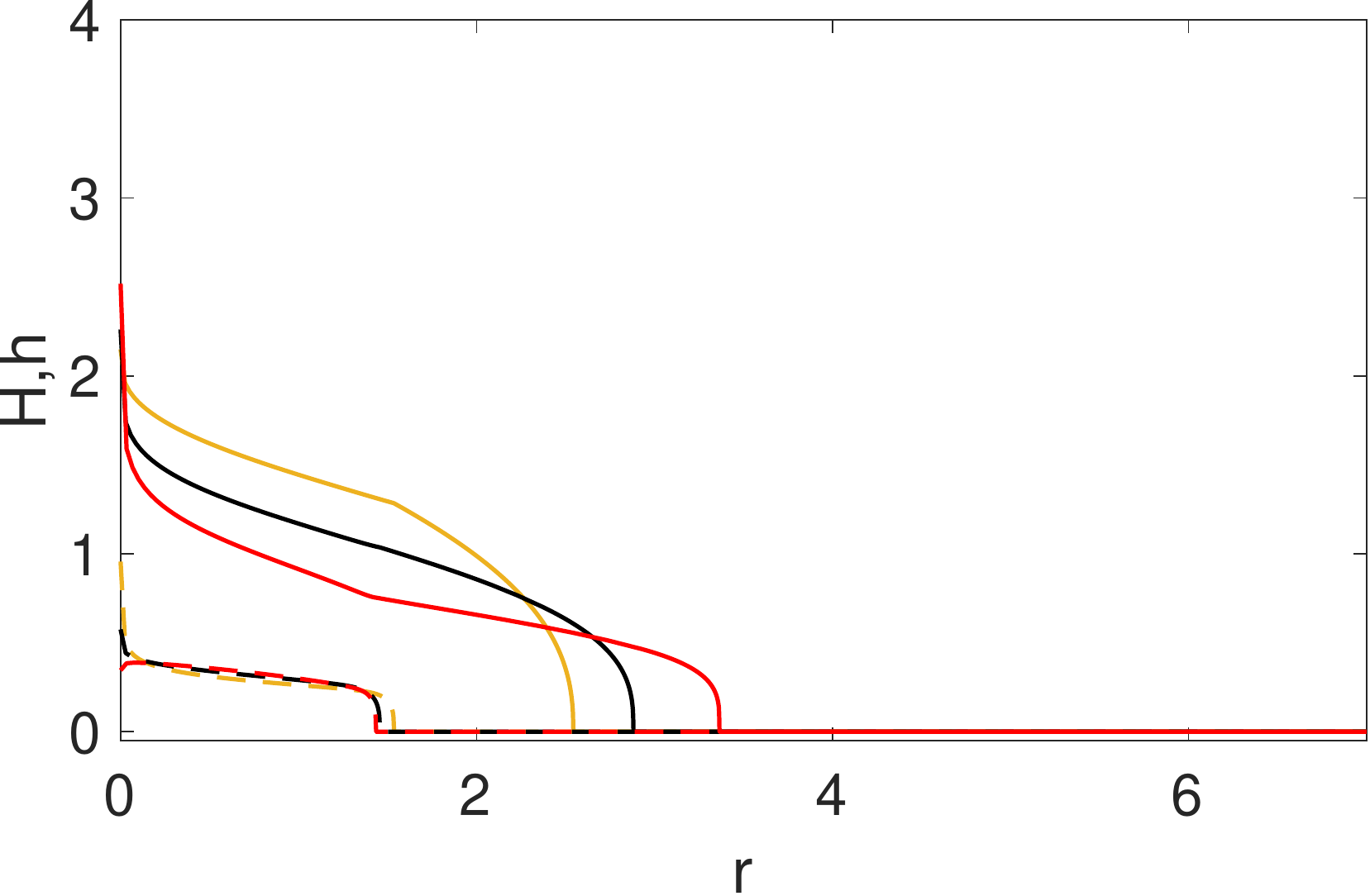}
\end{subfigure}
\begin{subfigure}{.5\textwidth}
\centering
\includegraphics[scale=0.37]{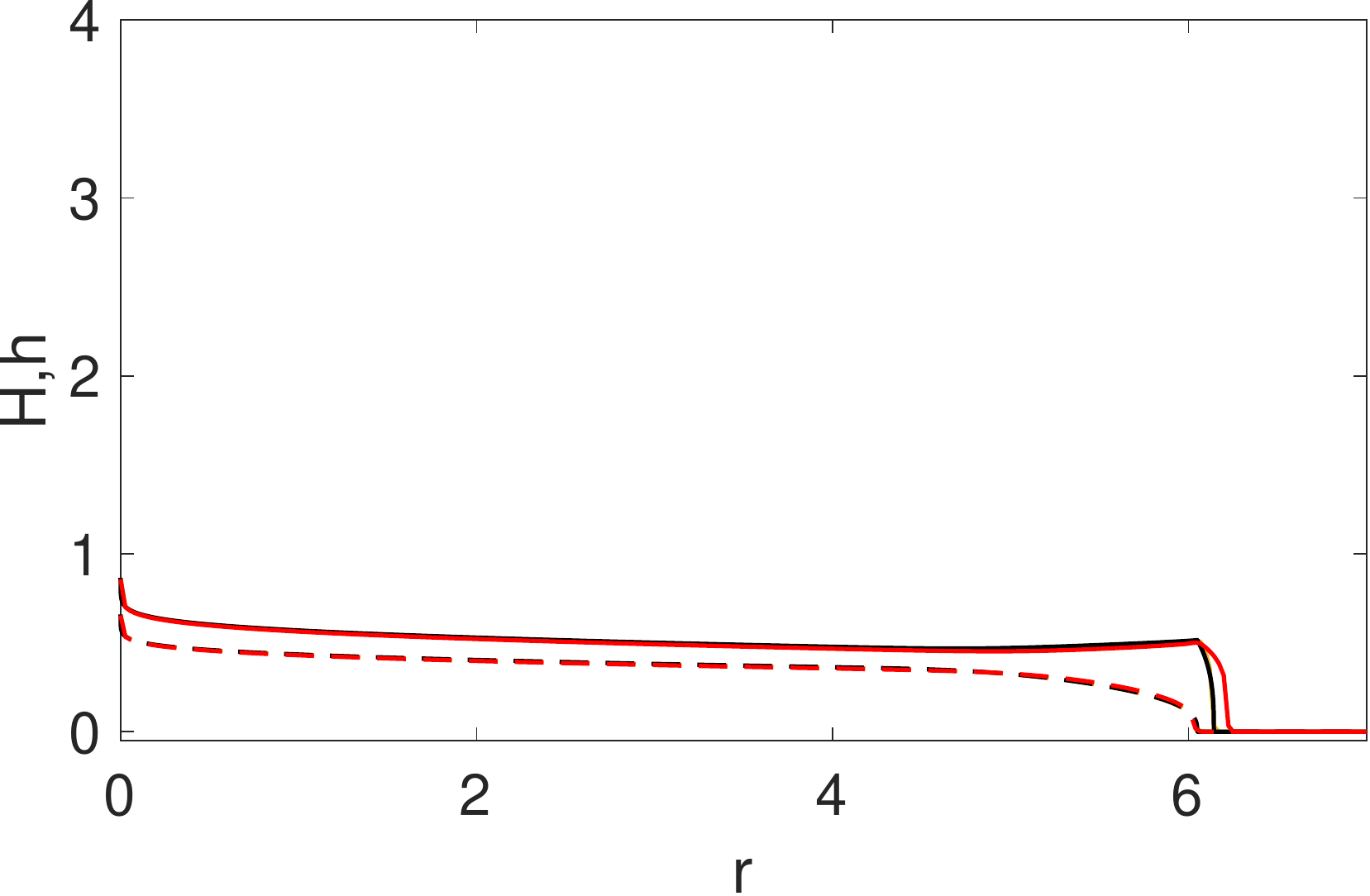}
\end{subfigure}
\begin{subfigure}{.5\textwidth}
\centering
\includegraphics[scale=0.37]{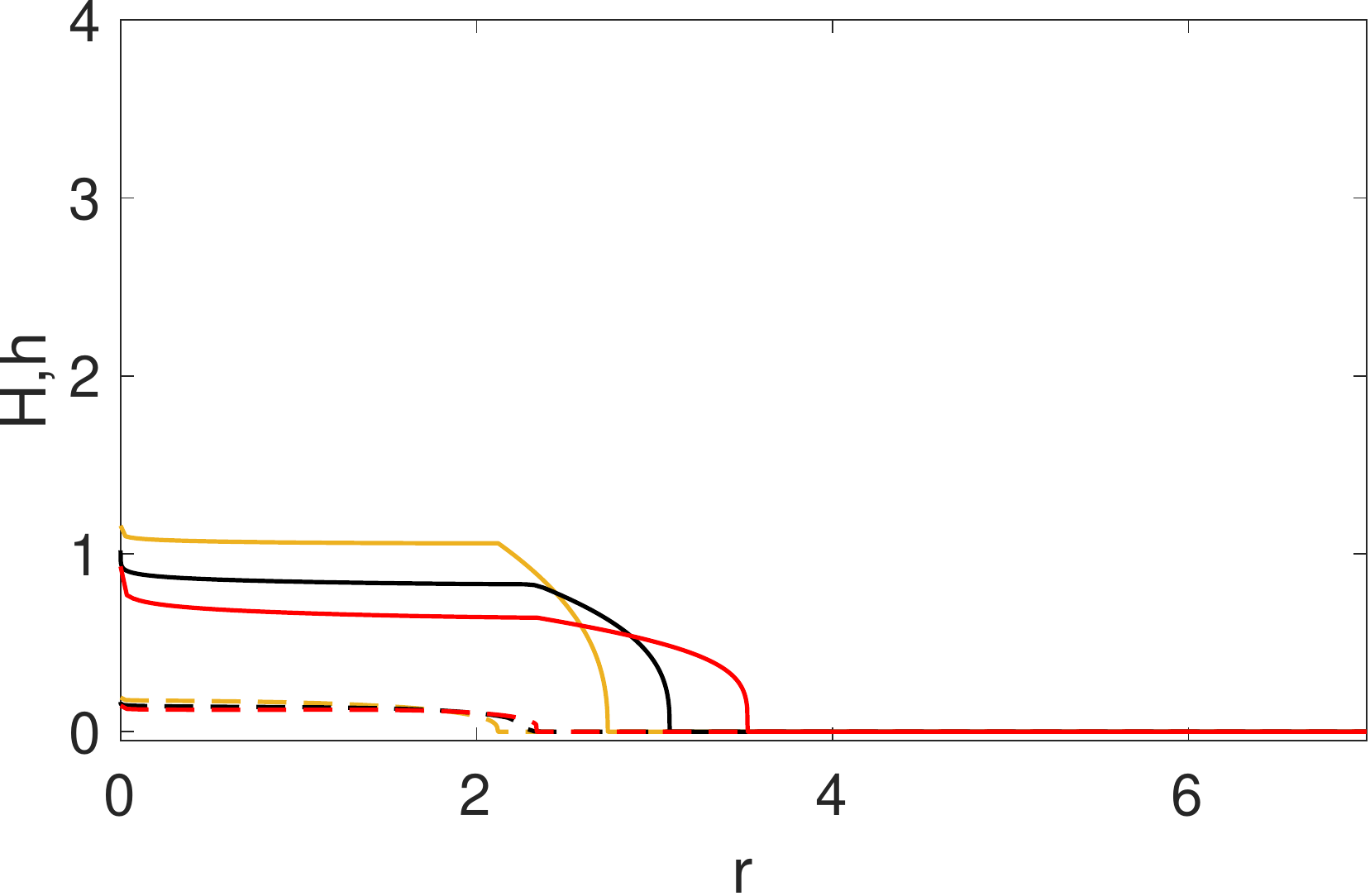}
\end{subfigure}
\begin{subfigure}{.5\textwidth}
\centering
\includegraphics[scale=0.37]{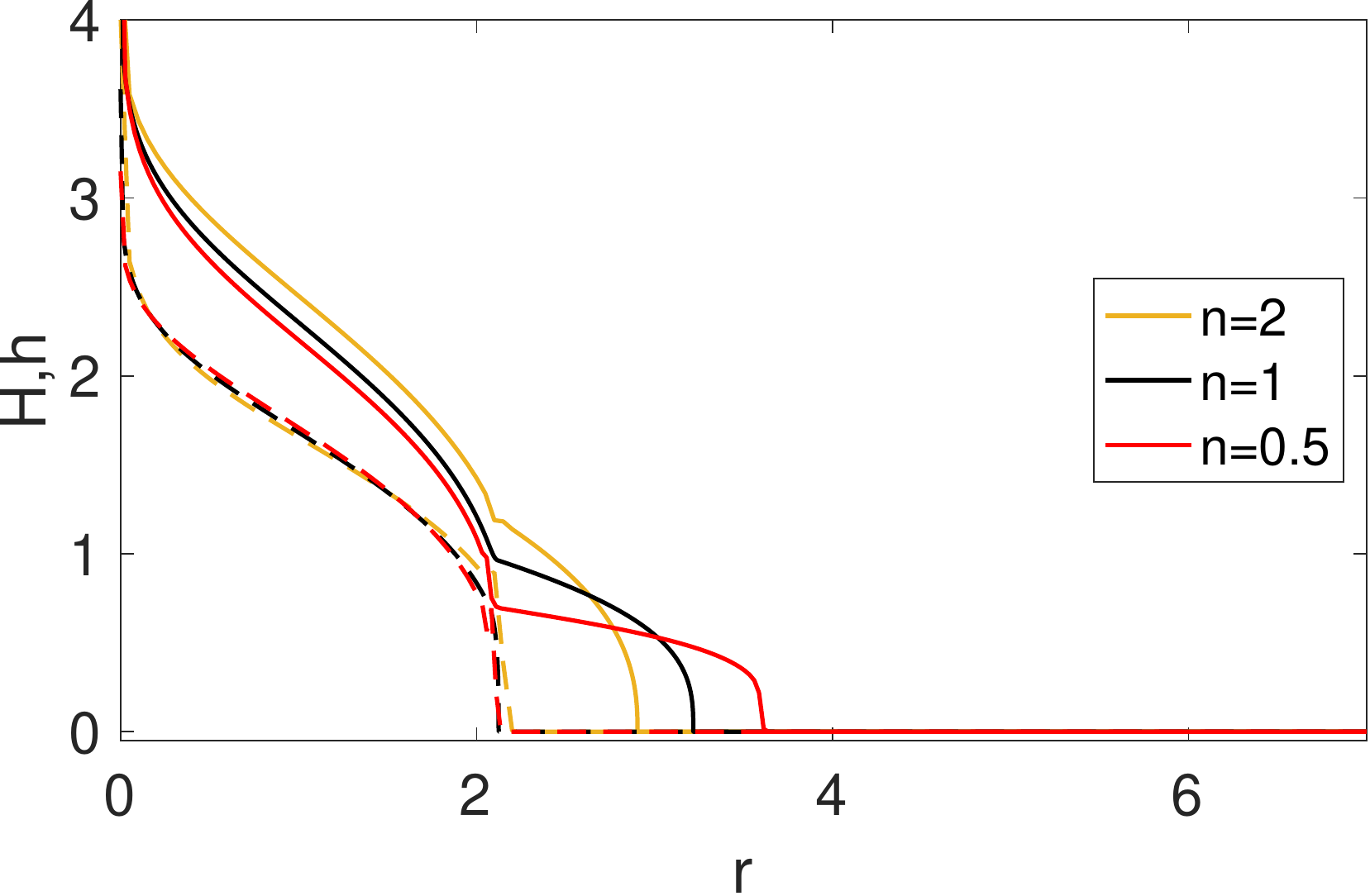}
\end{subfigure}  
\begin{picture}(0,0)(35,20)
  \put(100,240){\(I:\quad\M=1\),  \(\Q=0.1\)}
   \put(290,240){ \(II:\quad\M=100\),  \(\Q=2\)}
   \put(90,120){\(III:\quad\M=100\),  \(\Q=0.1\)}
   \put(290,120){\(IV:\quad\M=0.05\),  \(\Q=1\)}
\end{picture}
\caption{Characteristic solutions of each of the four states in the
  different regimes marked by $\blacklozenge$ in
  Figure \ref{fig:M-Q state map} for a range of power-law exponents (yellow
  $n=2$, black $n=1$ and red $n=0.5$), at constant flux
  (\(\alpha=1\)), showing the position of the free surface of the
  lubricated power-law fluid (solid), and the lubricating Newtonian
  fluid (dash), at constant dimensionless time t=20, and
  \(\mathscr{D}=1\). 
  \label{figure3}}
\end{figure}
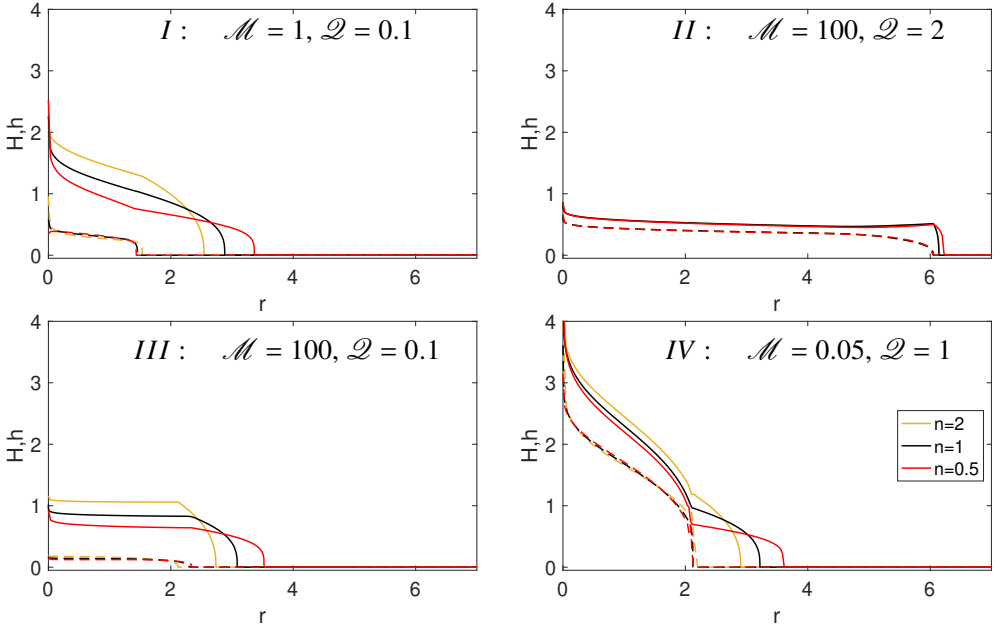

When the upper fluid is Newtonian $(n=1)$ the flow is self similar and
can be classified into four different flow regimes in the
\(\M\)-\(\Q\) state space \citep{KowalWorster:2015-JFM-Lubricated}, in
which the radial distribution of the fluid thickness and the evolution
of the fronts have unique qualitative characteristics (Figures
\ref{fig:M-Q state map}, \ref{figure3}).  When the upper fluid is
non-Newtonian (\(n \neq 1\)), no global similarity solution arises for
a constant-flux discharge (\S \ref{dimensionless}). Nevertheless, we
find numerically that the characteristic distributions of the fluid
thickness in each of the four regimes are qualitatively preserved
(Figure \ref{figure3}). Moreover, the thickness of the lubricating
fluid in the non-Newtonian case varies very weakly from that in the
purely Newtonian case, and the fronts $r_L$ in both cases appear to
follow a similar evolution pattern (Figure \ref{figure3}).
In contrast, the propagation rate of the front $r_N$ varies strongly
with the non-Newtonian properties of the upper fluid, becoming faster
than the purely Newtonian case for shear-thickening fluids and slower
than the purely Newtonian case for shear-thinning fluids (Figure
\ref{figure3}).  This pattern in all of the four regimes can be
physically rationalised through a dimension-based argument: the radial
velocity at constant flux $[u]\sim Q/(2\pi R H)$ combined with the
thickness scale $[H]\sim (Q\mu_\l/\rho g)^{1/4}$
(\ref{eq:TRHnewScales}c) implies that the dominant strain rate
$[\p u/\p z]\sim Q/(2\pi R H^2)\sim (\rho g Q/\mu_\l)^{1/2}/(2\pi R)$
declines radially. Consequently, a shear-thinning fluid (\(n > 1\))
becomes increasingly more viscous with radius than a shear-thickening
fluid, leading to its relatively slower propagation.

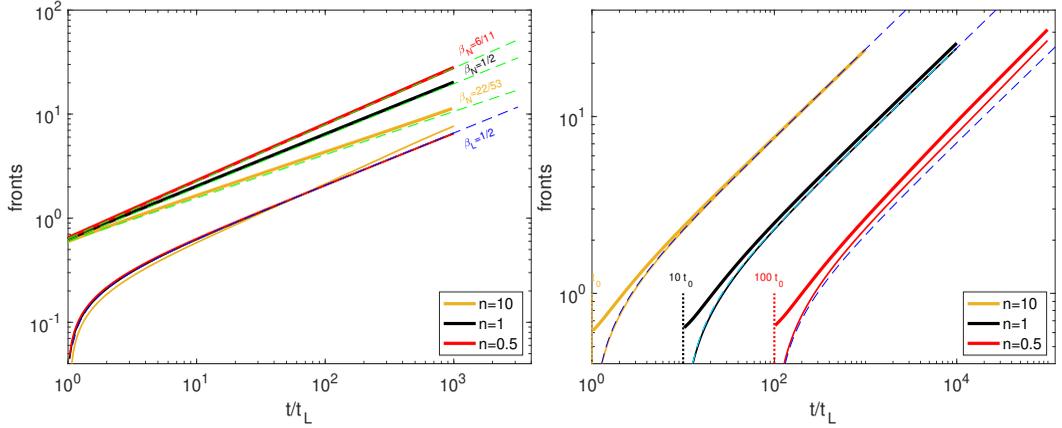
\begin{figure}
  \includegraphics[scale=0.4]{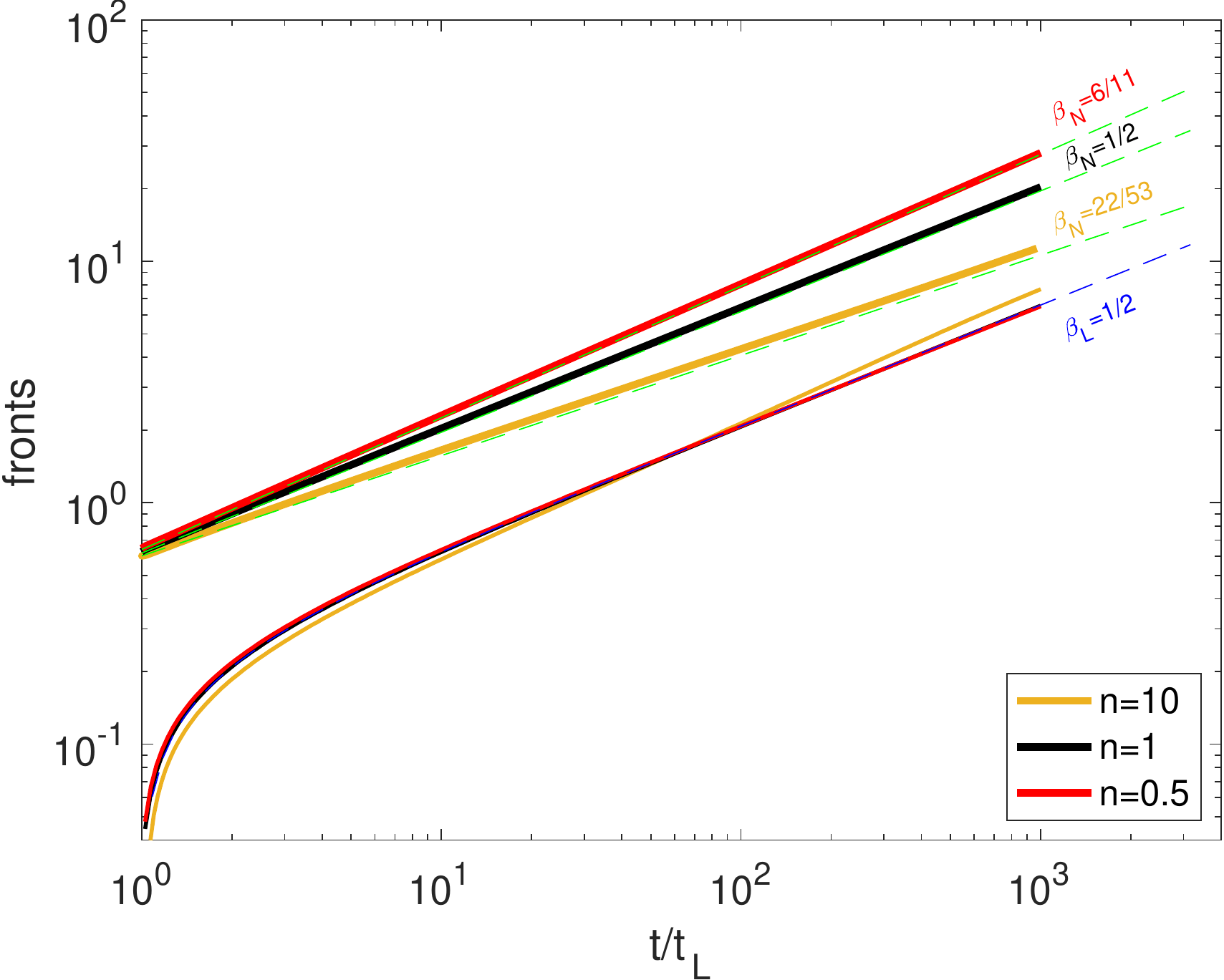}
  \includegraphics[scale=0.4]{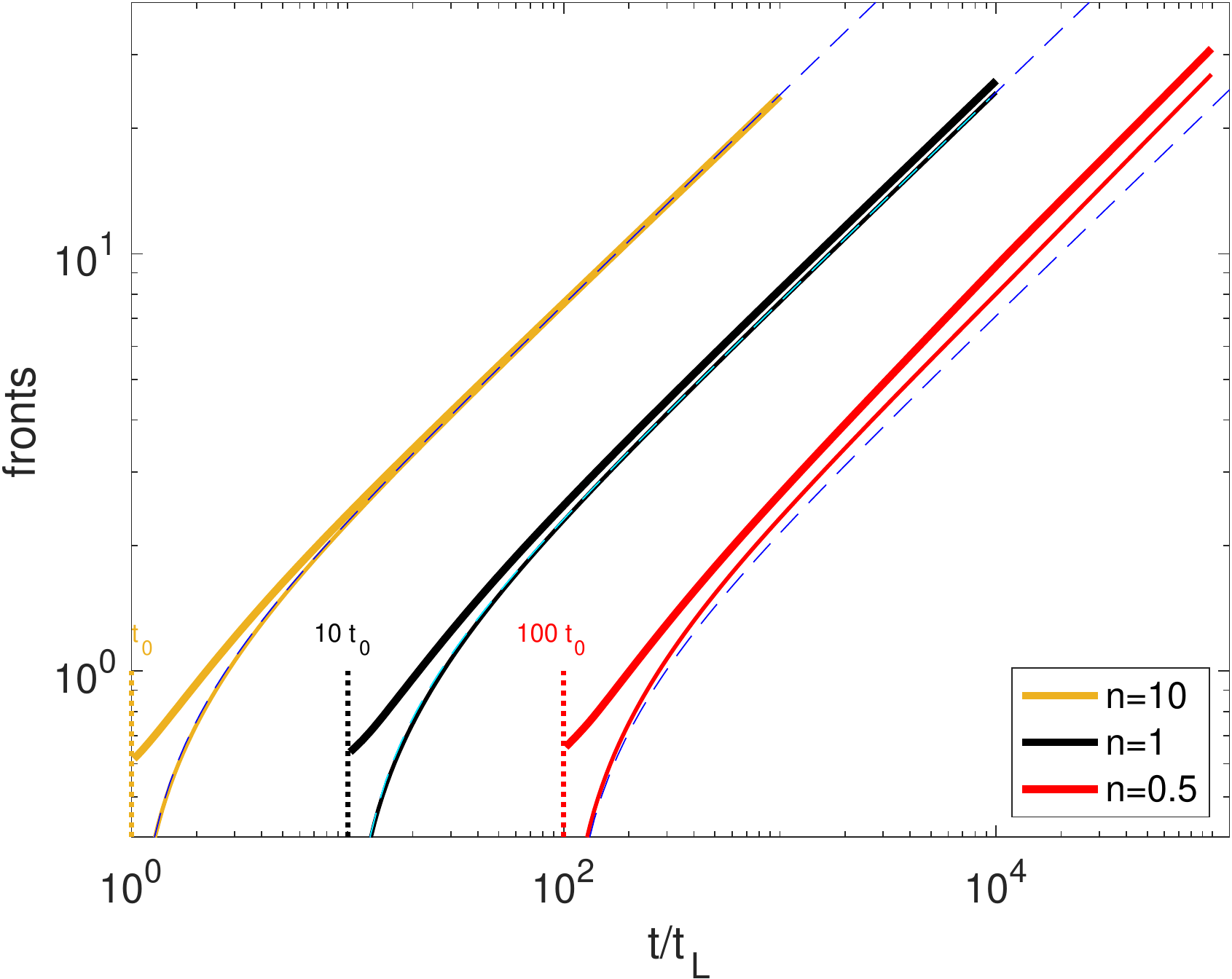}
\begin{picture}(0,0)(300,0)
   \put(00,100){$r_N$}
   \put(20,65){$r_L$}
\end{picture}
\caption{(a) The non-dimensional front positions $r_N$ (---, thick) and $r_L$ (---,
  thin) for  varying power-law fluid exponents, \(n=10,1,0.5\), and for
  dimensionless parameters \(\a=1\),
  \(\Q=0.2\) and \(\mathscr{D}=0.1\). (a)  \(\M=0.01\). 
  Shown for reference the theoretical prediction for $r_N(t)$ (eq. \ref{simMll}) in the
  \(\M\ll 1\) limit (- - -, green), and with the prediction for the purely
  Newtonian solution (- - -, blue).
  %
  (b) \(\M=10^4>\M_c(n)\). Shown for reference curves with exponent 1/2 (---, blue),
   corresponding to the purely Newtonian solution.
   The exponents of each front are $\beta_N=0.4962\pm 0.0006, 0.5\pm 
   0.001, 0.523\pm 0.003$ and  $\beta_L=0.4984\pm 0.0001, 0.4998\pm
   0.0003, 0.526\pm 0.002$ for \(n=10,1,0.5\), respectively.
   \label{frNLMsl}}
\end{figure}

\begin{figure}
  \includegraphics[scale=0.4]{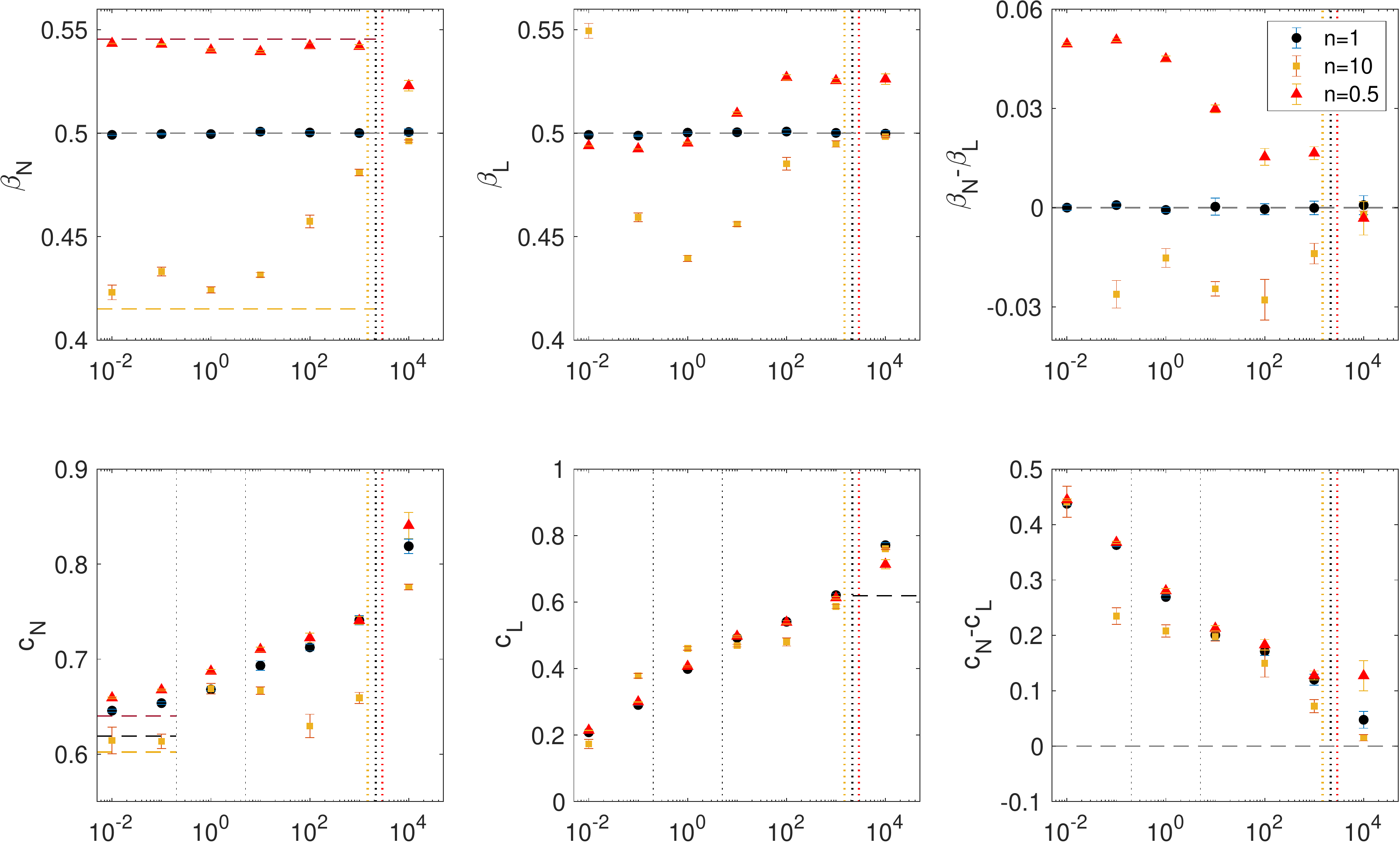}
  \begin{picture}(0,0)(375,15)
    \put(60,127){$\M$}
    \put(185,127){$\M$}
    \put(310,127){$\M$}
    \put(60,8){$\M$}
    \put(185,8){$\M$}
    \put(310,8){$\M$}
    \put(0,126){(a)}
    \put(130,126){(b)}
    \put(255,126){(c)}
    \put(0,7){(d)}
    \put(130,7){(e)}
    \put(255,7){(f)}
    \put(30,150){\small \textcolor{orange}{$\beta_N=22/53$}}
   \put(30,190){\small $\beta_N=1/2$}
   \put(30,229){\small \textcolor{red}{$\beta_N=6/11$}}
   \put(30,90){$IV$}
   \put(55,90){$I$}
   \put(75,90){$III$}
   \put(155,90){$IV$}
   \put(180,90){$I$}
   \put(200,90){$III$}
   \put(280,45){$IV$}
   \put(305,45){$I$}
   \put(325,45){$III$}
 \end{picture}
 \vspace{5mm}
 \caption{The fitted exponents and intercepts to numerical solutions
   of the fronts \(r_N,r_L\) for varying \(\M\), and \(n=10,1\) and
   \(0.5\), for \(\alpha=1\), \(\Q=0.2\) and
   \(\mathscr{D}=0.1\). Vertical grid line (dotted blue) marks the
   intercept-driven outstripping threshold $\M_c(n)$. (a) The
   exponent \(\beta_N\). Horizontal grid lines show the predicted
   asymptotic (\(\M\ll 1\)) exponents
   \((2n+2)/(5n+3)\)(eq. \ref{simMll}) corresponding by color to the
   specific \(n\)s.  (b) The exponent $\beta_L$. Horizontal grid line
   show the predicted asymptotic (\(n=1\)) exponent \(0.5\)
   \citep{KowalWorster:2015-JFM-Lubricated}.  (c) The exponent
   difference \(\beta_N-\beta_L\). The differences in $\M=10^4$ are
   $-0.002\pm 0.001, 0.0008\pm 0.003$, and $-0.003\pm 0.005$ for
   $n=10,1,0.5$, respectively.  Horizontal grid lines represents the
   critical value for exponent outstripping. (d) The intercept
   \(c_N\). Vertical grid lines (dash grey) mark the regime thresholds
   (Figure \ref{fig:M-Q state map}). Horizontal grid lines show the
   predicted asymptotic (\(\M\ll 1\)) intercepts
   $\xi_N\mathscr{N}^{1/(5n+3)}$ (eq. \ref{simMll}) corresponding by
   color to the specific \(n\)s. (e) The intercept $c_L$. Vertical
   grid lines (dash grey) mark the regime thresholds (Figure
   \ref{fig:M-Q state map}). Dashed grey curves are the theoretical
   prediction of \(c_L\) for \(n=1\)
   \citep{KowalWorster:2015-JFM-Lubricated}.  Horizontal grid line
   shows the asymptotic intercept of a non-lubricated Newtonian
   GC $\xi_N(1/3)^{1/8}$
   \citep{Huppert:1982-JFM-Propagation}. (e) The intercept difference
   \(c_N-c_L\). Horizontal grid lines mark the critical value for
   intercept outstripping.  
   \label{fig:betaNL_cNL}}
\end{figure}
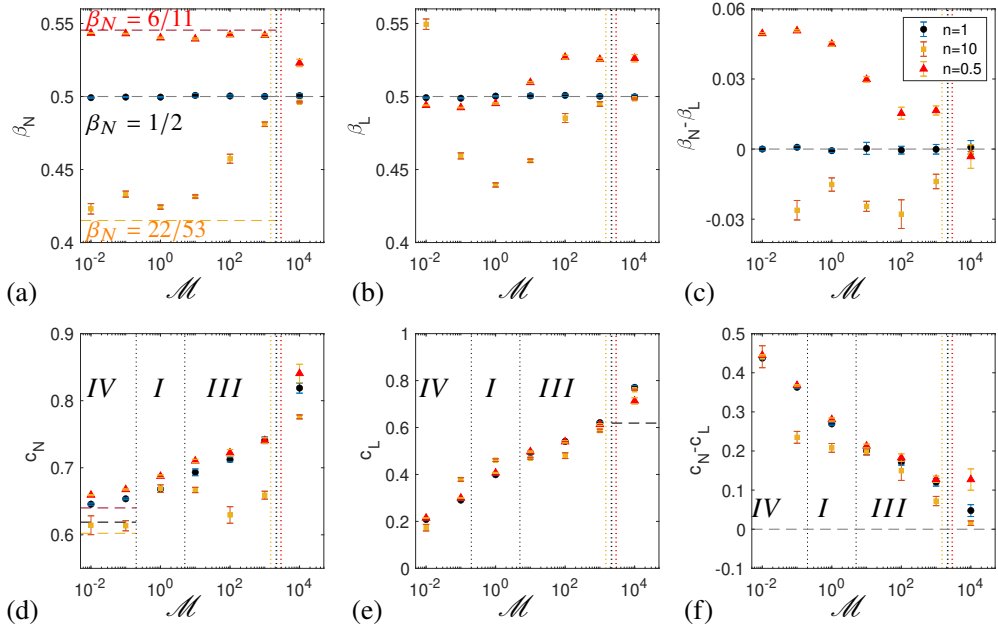

The slower front speed of a shear-thinning fluid compared to a
Newtonian fluid implies that outstripping by the front of the
Newtonian lubricating fluid may occur after some time
(\S\ref{sec:outstripping}). We investigate this more carefully by
considering the evolution of the two fronts within a range of
viscosity ratios $10^{-2}\le \M\le 10^4$, a range of fluid exponent
$n=10,1,0.5$ and for fixed flux ratio $\Q=0.2$ and density ratio
$\mathscr{D}=0.1$. Even though no global similarity solution of the
first kind exists when $\a,n\ne 1$, we find that each front has a
power-law evolution in time of the form
\begin{equation}
  \label{eq.form}
  r_N(t) \approx c_N t^{\beta_N}, \quad r_L(t) \approx c_L (t-1)^{\beta_L}
\end{equation}
in dimensionless units, where we compute the intercepts \(c_N, c_L\)
and the exponents \( \beta_N, \beta_L\), which may depend on the
dimensionless groups $\M,\Q,\mathscr{D}$ and $n$, through regression
to the numerical solution (Figures \ref{frNLMsl}, \ref{fig:betaNL_cNL}, and
\ref{fig:betaNL_cNL}).
Our numerical solver does not admit front outstripping since it
constrains \(r_L \leq r_N\). Therefore, we infer that
front outstripping is in progress when the interval \(r_N-r_L\) is
closing in time.

In the upper-fluid ``liquid'' limit, $\M \ll 1$, we find that the
numerical solution for the front of the upper fluid evolves
consistently with \eqref{simMll}, in which $\beta_N=(2+2n)/(5n+3)$ and
$c_N=\xi_N\mathscr{N}^{1/(5n+3)}$.
For example, when $\M=0.01$ the computed exponents $\beta_N$ for
$n=0.5,1,10$ differ from the corresponding theoretical values
$6/11, 1/2, 22/53$ by less than 2\% (Figure \ref{frNLMsl}a). The
larger discrepancy among those is for the shear-thinning fluid, which
is expected due to the stronger coupling in this case between the
upper and the lower fluid layers, as mentioned in
\S\ref{sec.invicid}. The corresponding intercepts $c_N$ differ by less
than 5\% from the theoretical values.
The lubricating fluid front \(r_L\) evolves with exponent
$\beta_L\approx 1/2$ for $n=1$, larger than $1/2$ for a shear-thinning
fluid (by 10\% for $n=10$), and smaller than 1/2 for shear-thickening
fluid (by 1.2\% for $n=1/2$) (Figure \ref{frNLMsl}a).  The corresponding
intercepts have small variation with $n$, being in the range
$0.17\lesssim c_L\lesssim 0.21$, which is much smaller than the
intercept in non-lubricated Newtonian GCs 
$\xi_L(1/3)^{1/8}\approx 0.62$.

In the upper-fluid ``solid'' limit, $M\gg 1$, we find that the front
of the lubricating fluid evolves consistently with \eqref{eq.Mgg1lim},
in which $\beta_L\rightarrow 1/2$ is $n$-independent, and the solution
for $\beta_N$ also tends to $1/2$ (Figure \ref{frNLMsl}b).
For intermediate \(\M\) values we find that the front exponent
\(\beta_N\) varies weakly with \(\M\) initially ($\M\lesssim 10$),
remaining close to the $\M\ll 1$ asymptotic values (Figure
\ref{fig:betaNL_cNL}a), and the intercept $c_N$ grows with $\M$ and varies
weakly with $n$ (Figure \ref{fig:betaNL_cNL}d).  The exponent of the
lubricating front $\beta_L$ grows with $\M$ over 1/2 for
shear-thickening fluids and less than 1/2 for shear-thinning fluids
(Figure \ref{fig:betaNL_cNL}b), whereas the intercept $c_L$ grows
monotonically with $\M$ for all $n$s while varying very weekly with
$n$ (Figure \ref{fig:betaNL_cNL}e).

Throughout the range $10^{-2}\le \M\le 10^4$ we find that
\(\beta_N > \beta_L\) for shear-thickening (\(n < 1\)),
\(\beta_N < \beta_L\) for shear-thinning fluids (\(n > 1\)), and
\(\beta_N = \beta_L=1/2\pm 0.0008\) when \(n=1\) (Figure
\ref{fig:betaNL_cNL}c), consistently with our predictions of the similarity
solutions near the front (\S\ref{sec:outstripping:exponent}).  This
implies that in the long-time limit the front $r_L$ outstrips $r_N$
when the upper fluid is shear thinning, independently of $\M$, while
no exponent outstripping is expected when the upper fluid is shear
thickening. However, the exponent differences
$|\Delta\beta|\equiv|\beta_L-\beta_N|$ declines with $\M$ so that when
$\M=10^4>\M_c$ it is marginally zero (Figure \ref{fig:betaNL_cNL}c),
implying termination of the exponent-driven
outstripping. Simultaneously, the intercept difference \(c_N - c_L\)
is positive throughout the range of $\M$ but approaches zero at the
vicinity of $\M=10^4$ (Figure \ref{fig:betaNL_cNL}f). This may reflect the
approach to an intercept-driven outstripping ($c_N-c_L<0$) across a
critical viscosity ratio $\M>\M_c$, given by \eqref{eq:Mc a=1}, while
both exponents converge to $1/2$. We note again that numerically we do
not resolve explicit outstripping, so the behaviour as $\M$ approaches
$\M_c$ and grows beyond it reflects mostly the qualitative trend.

\begin{figure}
\centering
\includegraphics[scale=0.36]{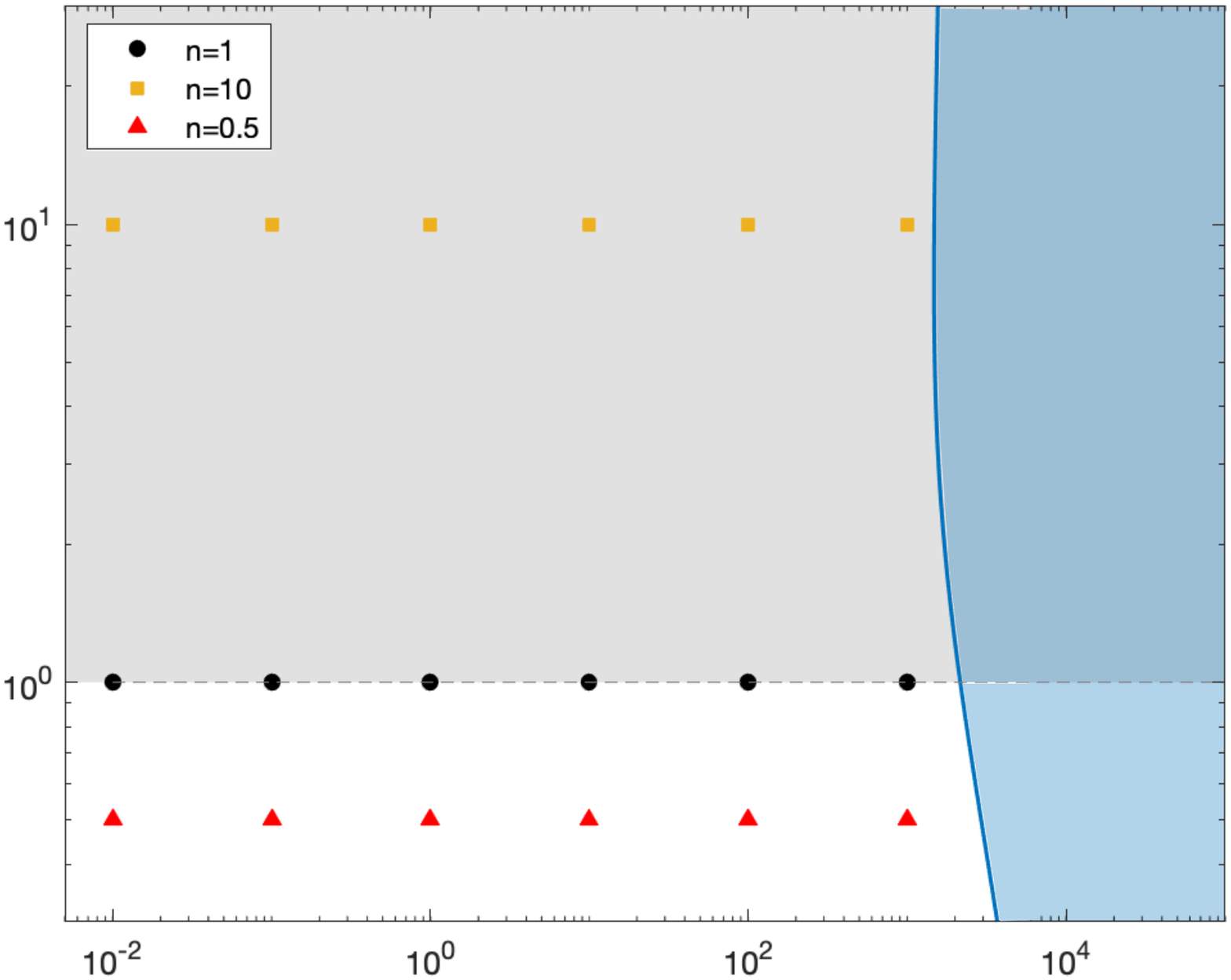}
\begin{picture}(0,0)(20,20)
  \put(-270,140){ $n$}
  \put(-200,165){\small Exponent-driven}
  \put(-190,155){\small outstripping}
  \put(-70,165){\small Intercept-driven}
  \put(-61,155){\small outstripping}
  \put(-285,165){\small shear-}
  \put(-290,155){\small thinning}
  \put(-285,65){\small shear-}
  \put(-295,55){\small thickening}
  \put(-135,15){$\M$}
  \put(-83,110){\rotatebox{95}{$\M_c(n)$}}
\end{picture}
\caption{The \(\M-n\) state map for $\a=1, \Q=0.2$, and $\D=0.1$,
  showing the numerical simulations presented in Figure
  \ref{fig:betaNL_cNL} (markers).  The region (cyan)
  bounded by the curve \(\M_c(n)\) is where Intercept-driven
  outstripping occurs. The region (gray) bounded by $n=1$ is where
  exponent-driven outstripping occurs.
  \label{fig.outstrippingMap}}
\end{figure}

\section{Comparison with experimental evidence for $\a=1$ and
  $n>1$} \label{sec:comparison}
  
The theory we have developed can be validated with the recent
laboratory experiments of lubricated GCs at constant-flux
($\alpha=1$) that consisted of a strain-rate softening fluid that was
lubricated by a denser Newtonian fluid
\citep{KumarZuriKoganEtAl2021JoFMLubricated}.  The experimental
findings have shown that for flux ratio $\Q\lesssim 0.06$ both fronts
were highly axisymmetric, in experiments that lasted over $10 t_L$ in
some cases, implying that a comparison with our axisymmetric theory
should be valid.

\subsection{Proparation of the fronts}

Experimentally it was found that the two fronts evolved faster than
those of non-lubricated GCs of the corresponding
fluids. Nevertheless, each front had a power-law time evolution with a
similar exponent as non-lubricated GCs of the
corresponding fluids. In particular, the front $r_N$ evolved with
exponent $(2n+2)/(5n+3)$ before the lubrication fluid was introduced
($t\le t_L$) as well as when $t\gtrsim 5t_L$, while the exponent was
larger during the transition between the two intervals. At the same
time, the front $r_L$ evolved with exponent 1/2.

\begin{figure} 
    \includegraphics[scale=0.45]{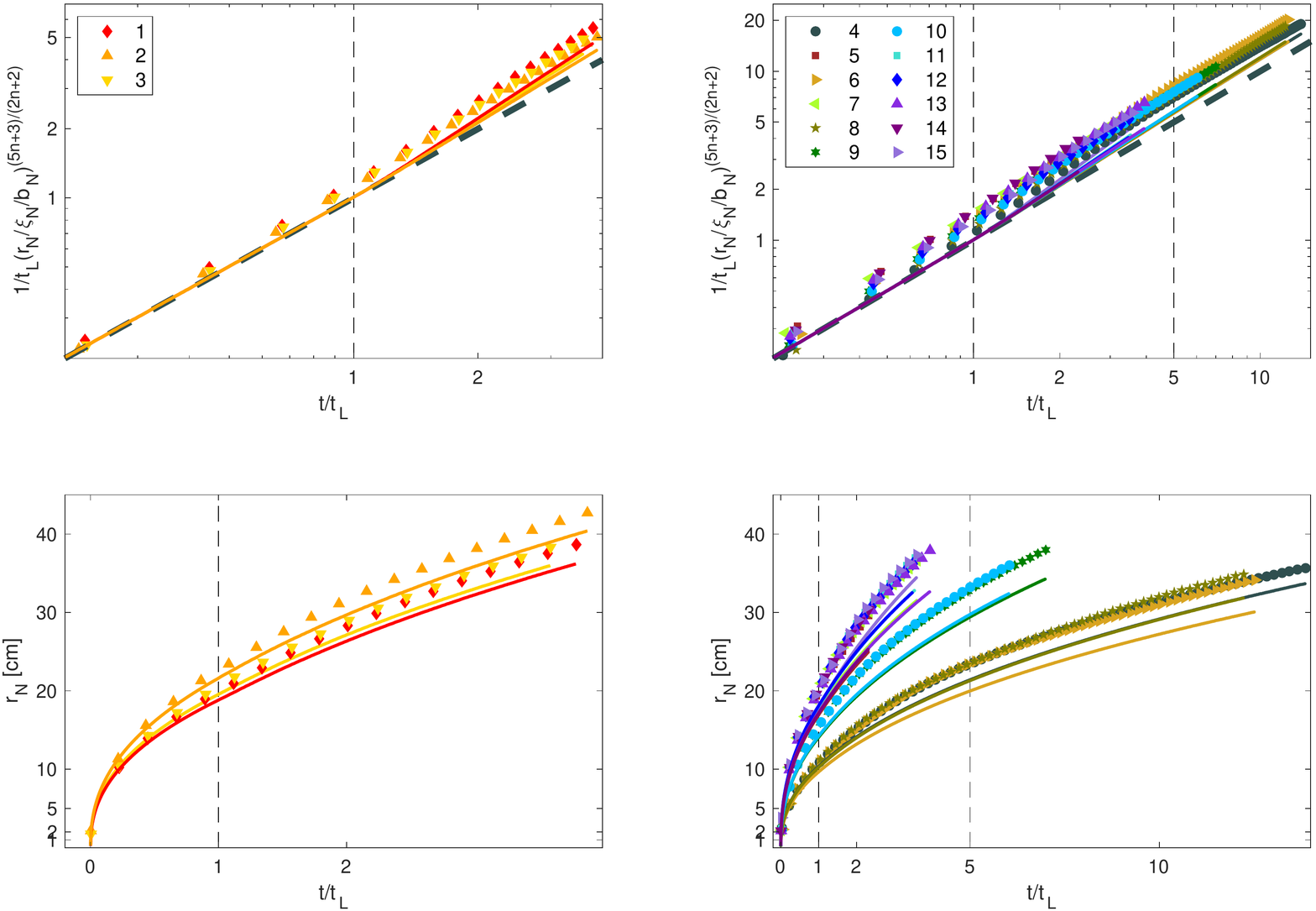} 
  \begin{picture}(0,0)(380,-145)
    \put(0,0){$(a)$}
    \put(200,0){$(b)$}
    \put(0,-145){$(c)$}
    \put(200,-145){$(d)$}
  \end{picture}  
\caption{Comparison of the theoretical predictions of the fronts $r_N$
  evolution (colored lines) with the experimental measurements
  (markers) \citep[][experiments  \#1-15 in table
  2c]{KumarZuriKoganEtAl2021JoFMLubricated}. (a, c) Comparison with the
  1\% polymer concentration (experiments 1-3) in linear and log scale
  respectively. In panel (a) the fronts are normalised with the
  similarity solution 
  of non-lubricated GCs of power-law fluids
  \citep{Sayag-Worster:2013-Axisymmetric} and the front solution of
  such curent is shown for reference (dash, gray). (b, d) Same as
  panels a and c but for experiments (4-15) of the 2\% polymer concentration.
  \label{fig:f4d:rN}}
\end{figure}

\begin{figure}
    \includegraphics[scale=0.45]{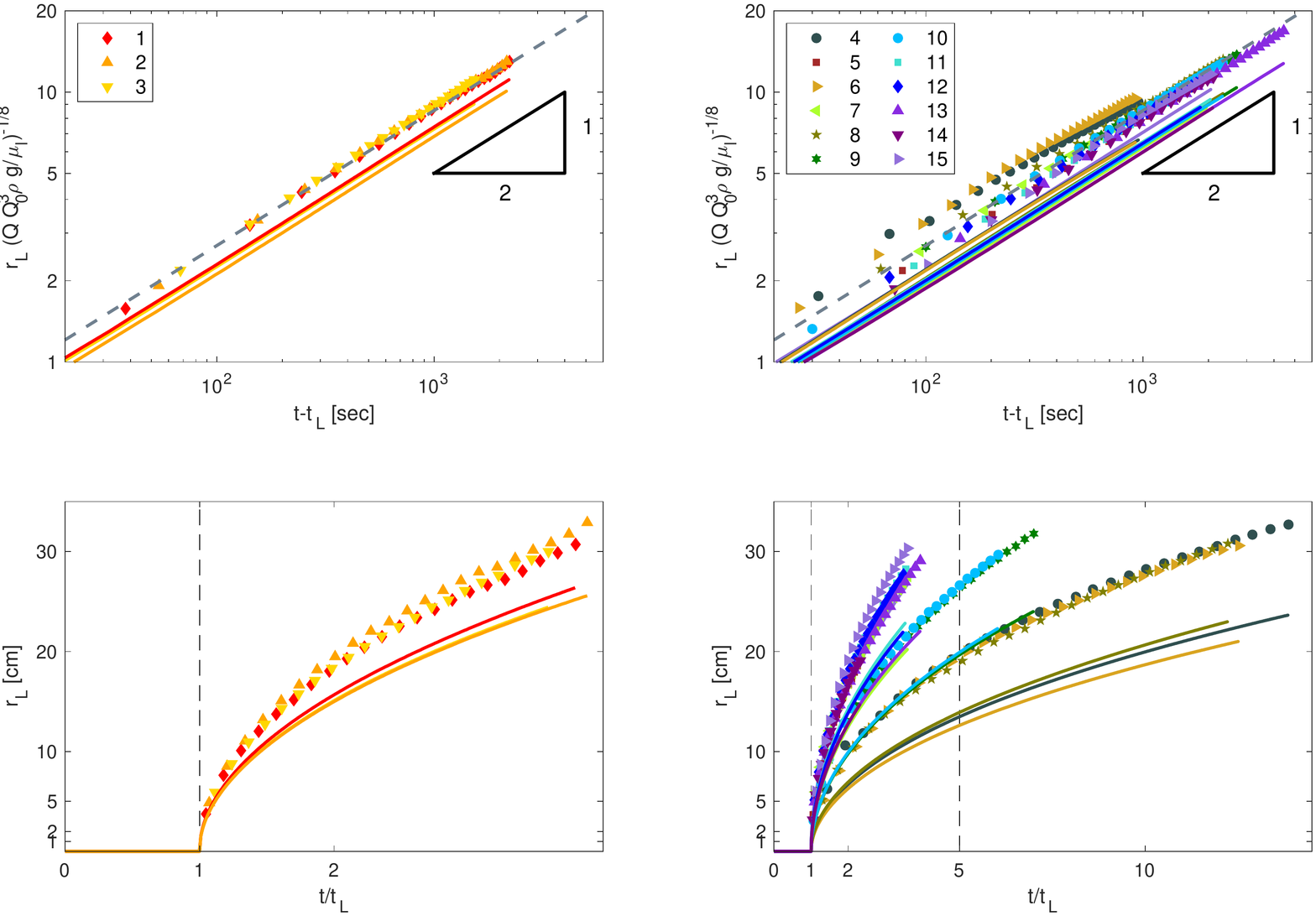} 
  \begin{picture}(0,0)(380,-145)
    \put(0,0){$(a)$} 
    \put(200,0){$(b)$}
    \put(0,-145){$(c)$}
    \put(200,-145){$(d)$}
\end{picture}
\caption{Comparison of the theoretical predictions of the fronts $r_L$
  evolution (colored lines) with the experimental measurements
  (markers) \citep[][experiments \#1-15 in table
  2c]{KumarZuriKoganEtAl2021JoFMLubricated}. (a-b) Comparison with the
  1\% polymer concentration (experiments 1-3) in linear and log scale
  respectively. In panel (b) the fronts are normalised with the
  similarity solution of a Newtonian lubricated GCs
  \citep{KowalWorster:2015-JFM-Lubricated} and the front solution of
  such a current is shown for reference (dash, gray), with a fitted
  coefficient 0.27. (c-d) Same as
  panels a and b but for experiments (4-15) of the 2\% polymer
  concentration.
  \label{fig:f4d:rL}}
\end{figure}

To compare our theoretical predictions to the experimental
measurements we compute numerical solutions to each of the 15
experiments described in \citet[][Table 2c and
eq. 2.2]{KumarZuriKoganEtAl2021JoFMLubricated} (Figures
\ref{fig:f4d:rN}, \ref{fig:f4d:rL}) based on the measured quantities
$\Q,\M,\D$ and $n$ and without any fitting parameter. The viscosity
ratio in each of these experiments was in the range $1 \ll \M<\M_c$,
implying that the experiments were all in the exponent-outstripping
regime (\S\ref{sec:outstripping:exponent}) as well as in the
solid-limit regime (\S\ref{sec.icesheet}). Consequently, we expect the
lubrication front to evolve like $r_L\propto t^{1/2}$.

Comparing the time exponents, we find that the numerical solutions and
the theoretical predictions are consistent with those measured
experimentally for both $r_N$ (Figure \ref{fig:f4d:rN}a, b) and $r_L$
(Figure \ref{fig:f4d:rL}a, b).  Specifically, during
$t_L<t\lesssim 5t_L$ the solutions to the front $r_N$ evolve with
larger exponent than non-lubricated current (Figure
\ref{fig:f4d:rN}a,b), and when $t\gtrsim 5t_L$ the exponent diminishes
and converges back to that of non-lubricated GCs (Figure
\ref{fig:f4d:rN}b). Similarly, the solutions to the front $r_L$ evolve
with an exponent 1/2, consistently with the experimental measurements
(Figure \ref{fig:f4d:rL}a, b). In spite of the exponent consistency
between the experiment and the theory, the experimental fronts advance
faster than the predicted ones (Figure \ref{fig:f4d:rN}c, d and
\ref{fig:f4d:rL}c, d). This discrepancy is larger in experiments with
higher polymer concentration (Figure \ref{fig:f4d:rN}d and
\ref{fig:f4d:rL}d). It implies that the numerically predicted
intercepts are lower than those that were measured experimentally,
suggesting that additional physical processes that contribute to a
faster front propagation are not accounted for by our theoretical
model. We elaborate on potential additional mechanisms in
\S\ref{sec:discussion}.
  
\subsection{Thickness evolution}
\label{sec:exp-thickness}

\begin{figure} 
      \includegraphics[scale=0.45]{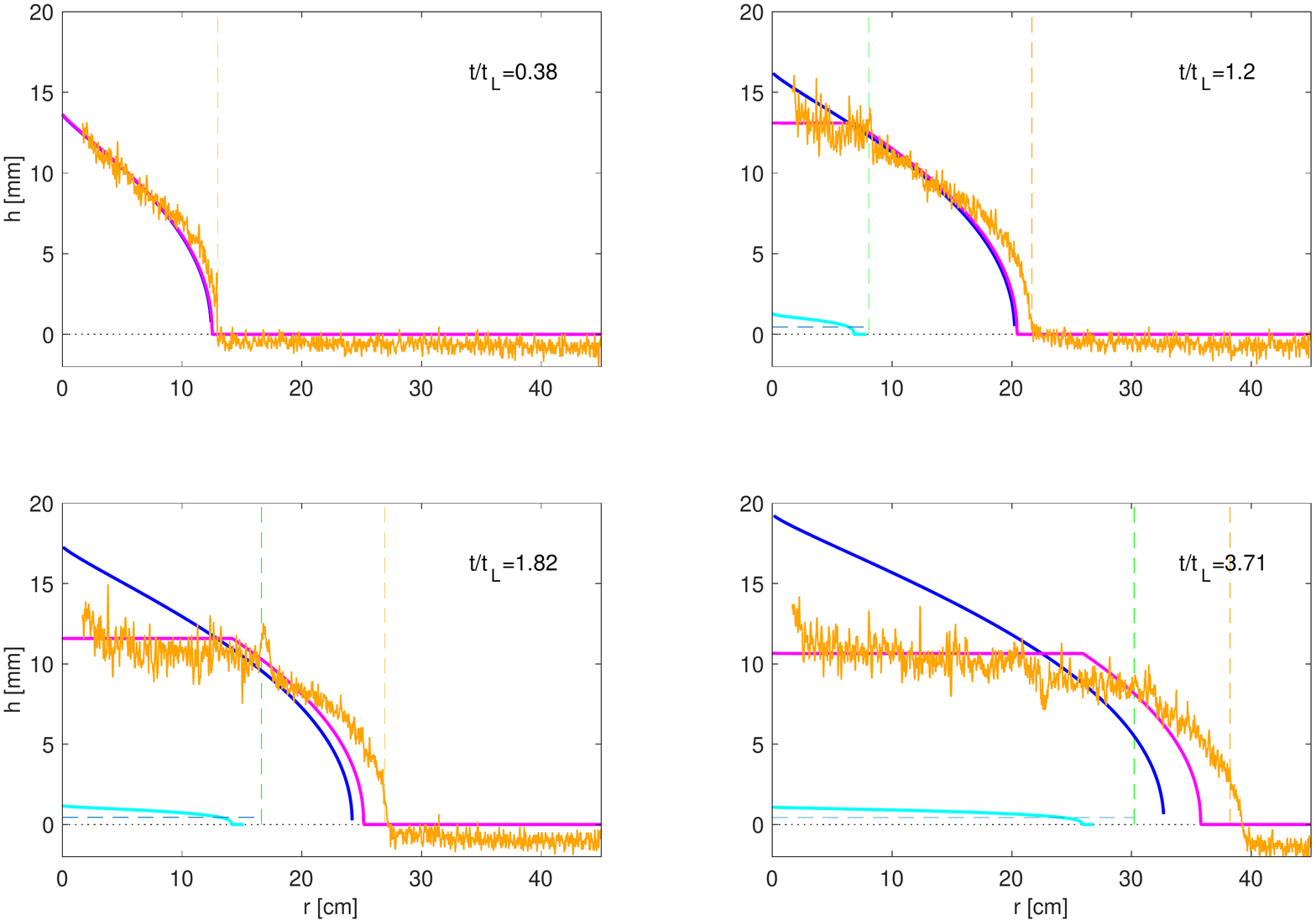} 
  \begin{picture}(0,0)(380,-145)
    \put(0,0){$(a)$}
    \put(200,0){$(b)$}
    \put(0,-145){$(c)$}
    \put(200,-145){$(d)$}
  \end{picture}  
  \caption{The thickness field of the top fluid layer along a radius at four different times during
    the non-lubricated phase (a) and during the lubricated phase
    (b-d). The experimental thickness measurement derived from
    transmitted-light intensity (---, orange)
    \citep[][experiment \#1 in table
    2c]{KumarZuriKoganEtAl2021JoFMLubricated} compared with the
    corresponding solution \citep{Sayag-Worster:2013-Axisymmetric} for
    the thickness of a non-lubricated GC of PL fluid (---, blue) and
    with the numerical solution for lubricated power-law fluid (---,
    magenta). Also shown are the corresponding numerical solution for
    the lubrication layer (---, cyan), and the experimentally measured
    fronts $r_N$ (vertical grid line, orange),  $r_L$ (vertical grid
    line, green), and the average thickness of the lubrication fluid
    $h_\l$ (- - -, pale blue).
    \label{fig:f6c:Hh}}
\end{figure}

Experimentally, the thickness of the lubricated, non-Newtonian fluid
was found to be nearly uniform in the lubricated region. The layer of
the lubricating fluid was also largely uniform with localised spikes,
and its average thickness was approximated through mass conservation
to be $h_\l=Q_\l(t-t_L)/\rho_\l\pi r_L^2$
\citep{KumarZuriKoganEtAl2021JoFMLubricated}. This nearly uniform
pattern differs substantially from the monotonically diminishing
thickness of a non-lubricated GC under similar
conditions. Our numerical solutions, which do not involve any fitting
parameter, follow a similar pattern as in the experiments (Figure
\ref{fig:f6c:Hh}). The theoretical and experimental patterns are
highly consistent during most of the flow (e.g., Figure
\ref{fig:f6c:Hh}a-c), but discrepancy between the two grows
progressively near the fronts $r_L, r_N$ and in the non-lubricated
region (e.g., Figure \ref{fig:f6c:Hh}d). In particular, the fronts in
the experiment evolve faster than those computed numerically, and the
thickness distribution of the top fluid at the vicinity of the
lubricating front changes more sharply in the computed solutions than
in the measurements. This growing discrepancy may imply the action of
additional mechanisms, as we elaborate in \S\ref{sec:discussion}.

\section{Discussion} \label{sec:discussion}

The lubricated GCs that we consider involve five
dimensionless parameters $\mathscr{Q},\mathscr{D},\mathscr{M}, n$ and
$\alpha$, associated with significant qualitative transitions in the
structure of the solution, in the relative motion of the fluid fronts
and their stability, and in the thickness distribution of the two fluids.

The fluid exponent $n$ and the discharge exponent $\a$ have a dramatic
qualitative impact on the solutions. Specifically, similarity
solutions exist only when $n=1$, in which both fluids follow a similar
constitutive law, and when $\a=5$. In those cases the two fronts
evolve with the same power law in time and as a result the ratio
$r_N/r_L$ is constant. In all other cases ($n\ne 1,\a\ne 5$) the
fronts also appear to follow a power law evolution, but each with a
different exponent, implying that asymptotically in time the ratio
$r_N/r_L$ evolves following a power law in time with exponent
$\Delta\beta$ \eqref{eq:dphi}.  Therefore, there are solutions where
$r_N/r_L$ declines in time resulting in the lubrication front
outstripping the front of the upper fluid ($n>1~\&~\a<5$, and
$n<1~\&~\a>5$), and otherwise $r_N/r_L$ grows in time (Figure
\ref{figure_outstrip_exp}).

The flux ratio $\mathscr{Q}$ can lead to the emergence of two
significantly different patterns, and may have a critical impact on
the front stability. When $\Q<1$ the flux of the lubricating fluid is
lower than the top fluid layer. Consequently, the thickness of the
lubricating layer is significantly smaller and the propagation of the
front $r_L$ is affected by the relatively larger pressure imposed by
the thicker top layer (Figure \ref{figure3}III, and Figure
\ref{fig:f6c:Hh}).  The opposite occurs when $\Q>1$ -- the lubricating
fluid is discharged at a larger flux and its thickness is
significantly larger than the top fluid layer (Figure
\ref{figure3}II).
$\Q$ may also have a crucial impact on the stability of the
axisymmetric fronts.  Preliminary experimental evidence indicate that
when the flux is constant ($\a=1$), the top fluid layer is strain-rate
softening ($n=6$) and $\M\gg 1$, the initially axisymmetric fronts
become unstable when $\Q\gtrsim 0.1$ and develop fingering patterns
after an initial axisymmetric spreading
\citep{KumarZuriKoganEtAl2021JoFMLubricated}. Similar symmetry
breaking also emerges in the purely Newtonian case when
$0.14\lesssim \mathscr{Q}\lesssim 0.44$
\citep{KowalWorster:2015-JFM-Lubricated}.

The viscosity ratio $\M$ affects the relative motion of the fronts,
and the relative thickness of the fluid layers.  At a high viscosity
ratio ($\mathscr{M}\gg 1$) the more viscous top fluid is effectively
solid-like compared with the less viscous lower fluid. Consequently,
the flow in the lubricated region is independent of the fluid exponent
$n$, and both fluid layers in that region follow the same similarity
solution, in which the front of the lubrication fluid $r_L$ evolves
 with a time exponent $(3\a+1)/8$, same as a non-lubricated
Newtonian GC \eqref{eq.Mgg1lim}.
In the low viscosity ratio ($\mathscr{M}\ll 1$) the top fluid is
significantly more mobile than the lower fluid layer, which does not
provide an effective lubrication.  In this case the two fluid layers
in the lubricated region do not exhibit a global similarity solution,
but the top fluid layer along the whole domain does.  Consequently, a
self-similar solution exist in the top-fluid layer, in which the front
$r_N$ evolves with a time exponent $[\a(2n+1)+1]/(5n+3)$, same as a
non-lubricated GCs \eqref{simMll}.
The impact of the viscosity ratio on the fluid thickness distributions
can be appreciated through the constant flux case ($\a=1$), in which
the free surface of the top fluid is substantially flatter in the
$\M\gg 1$ case than in the $\M\lesssim 1 $ case (Figure
\ref{figure3}).


Independently of the value of the dimensionless parameters, the
solutions at the vicinity of the fronts are also self similar, with
exponents consistent with those of non-lubricated GCs of
power-law fluids
\citep{Huppert:1982-JFM-Propagation,Sayag-Worster:2013-Axisymmetric}. Particularly,
both fronts evolve with an exponent $[\a(2n+1)+1]/(5n+3)$, which
simplifies to $(3\a+1)/8$ for the Newtonian lubricating fluid. The
intercepts of the fronts depend on the different dimensionless numbers
of the system. Together, the exponents and the intercepts provide
insights into the interaction between the two fronts.  One important
consequence of that interaction is the outstripping of the upper fluid
front by the lower lubricating fluid front, which can occur either
through the intercept difference $\Delta\eta$ or through the exponent
difference $\Delta\beta$.
The condition for an intercept-driven outstripping can be formalised
in terms of the critical viscosity ratio $\M_c(\Q,\D,\a,n)$
\eqref{eq:Mc n=1}, so that outstripping occurs when
$\M>\M_c$. Physically this implies that when $\Q\ll 1$ then
$\M_c\propto 1/\Q^3 \gg 1$ and the top fluid should be significantly
more viscous than the lower fluid for outstripping to occur, in which
case the top fluid deforms substantially slower making it easier for
the lubricating fluid front to outstrip. Alternatively, when $\Q\gg 1$
then for shear-thinning fluids
$\M_c(n\rightarrow \infty)\propto \Q^{1/5}\gg 1$, whereas for
shear-thickening fluids
$\M_c(n\rightarrow 0)\propto 1/\Q^{1/3}\ll 1$. In either case the
superiority of the lubricating fluid flux leads to front outstripping.
When $\M<\M_c$ then $\Delta\eta >0$ and outstripping can only be
driven by the exponent difference. As discussed above, the condition
for that mechanism depends on the values of $n$ and $\a$ (Figure
\ref{figure_outstrip_exp}).  For example, shear-thinning fluids at
relatively low discharge exponent $(\a<5)$ become increasingly more
viscous as they expand radially, resulting in slower front velocity
than the lubricating fluid front. Moreover, their thickness and
correspondingly the pressure they apply on the lubricating fluid is
relatively larger and contribute further to the radial spreading of
the lubricating fluid.
We find that the intercept-driven outstripping can occur significantly
faster than exponent-driven outstripping. For example, as in the
constant flux ($\a=1$) case intercept-driven outstripping occurs at
roughly $t/t_L\lesssim 10$, which is much faster than the
\(t/t_L\gtrsim 10^3\) in the exponent-driven case (Figure
\ref{frNLMsl}).

It is important to note that the global similarity solution that we
find for a discharge exponent $\alpha=5$ arises in additional
axisymmetric GCs of different settings, which a priori
appear remotely related. This includes for example, isothermal lava
domes that are modeled as axisymmetric GCs of
visco-plastic fluids \citep[][]{Balmforth:2000-Visco-plastic}.  The
structure of the similarity solution in that case is identical to the
lubricating GCs that we consider, in which the front
evolves like $r\propto t^2$ and the fluid thickness evolves like
$h\propto t$ independently of the fluid exponent $n$.  Another system
with a similarity solution at $\alpha=5$ is the axisymmetric viscous
GCs flowing over a porous medium
\citep{Spannuth:2009-Axisymmetric}. Such a similarity among a broad
range of physical systems may not be coincidental and could imply a
more general symmetry associated with the circular geometry.

Many aspects of the theory were found consistent with experiments
performed for the dimensionless parameters
$\D\approx 0.15, \Q<0.06, 1\ll\M<\M_c, \a=1$, and $n>1$
\citep{KumarZuriKoganEtAl2021JoFMLubricated}. In particular, the time
evolution of both fluid fronts predicted by the theory is consistent
with the power law measured in the experiments. In addition, the
thickness distribution we predict for the top fluid layer is in good
agreement with the experimental measurements. However, some
discrepancies that arise may imply that the theory is not entirely
complete. Specifically, the theoretical predictions for the intercepts
do not accurately capture the measured ones, particularly in the case
of the lubricating front, which evolves faster than the theoretical
predictions.  Several potential physical mechanisms that the present
theory does not account for may explain these discrepancies.
One possible mechanism is that the lubrication front $r_L$ advances as
a hydrofracture in between the substrate and the relatively solid
viscous fluid layer \citep[][]{BallNeufeld:2018-PRF-Static}, or as a
shock in the fluid-fluid interface at $r_L$
\citep{DauckBoxGellEtAl:2019--Shock}.  In addition, discrepancy
between the experimental measurements of $r_N$ before introducing the
lubrication fluid ($t/t_L<1$) and the theoretical prediction of a
non-lubricated GC, particularly for the 2\% polymer
concentration \citep{KumarZuriKoganEtAl2021JoFMLubricated}, may imply
that the power-law constitutive equation that we use is incomplete.
Specifically, the time for the viscosity to adjust to the evolving
strain rates may not be instantaneous as we assume, but finite. In
addition polymer entanglements may arise in higher polymer
concentrations that potentially drive wall slip at the fluid-solid
interface through adhesive failure of the polymer chains at the solid
surface or through cohesive failure owing to disantanglement of chains
in the bulk from chains adsorbed at the wall
\citep{BrochardGennes:1992-L-Shear}. The implications of these
potential physical mechanisms will be addressed in future studies.

\section{Conclusions} \label{sec:conclusion}

Lubricated gravity currents are controlled by complex interactions
between two fluid layers. The lower lubricating layer modifies the
friction between the substrate and the top layer, which in turn
applies stresses that affect the distribution of the lubricating
layer. The resulting flow can vary dramatically from non-lubricated
gravity currents.

Unlike previous axisymmetric gravity current models that involve a
single fluid layer
\citep{Huppert:1982-JFM-Propagation,Sayag-Worster:2013-Axisymmetric},
or two coupled layers that have the same constitutive structure
\citep{KowalWorster:2015-JFM-Lubricated}, the flow we consider does
not in general admit a global self-similar solution. Exceptional
cases, in which the model has a global similarity solutions are the
purely Newtonian case and the case of a discharge exponent
$\a=5$. Several other situations admit a similarity solution in part
of the domain. This includes the asymptotic limits of the viscosity
ratio, corresponding to the top layer solid ($\M\gg 1$) and liquid
($\M\ll 1$) limits, and the solution at the vicinity of the fluid
fronts. The latter implies that the time evolution of each front is
proportional to that of the non-lubricated front of the corresponding
fluid. This implies that generally the ratio of the two fronts
positions $r_N/r_L$ either diverge in time or converge to zero, and
that the difference of the front intercepts can change
sign. Consequently, there are situations where the lubricating fluid
front can outstrip the outer-fluid front. This situation can arise
when the top fluid is non-Newtonian, having a different exponent than
the lubricating fluid, but also when there is global similarity and
the viscosity ratio is larger than a critical value $\M_c$ that
depends on the fluids flux and density ratios, on the discharge
exponent and on the power-law fluid exponent. %
In the canonical case of constant flux the flow we consider has no
global similarity when the top fluid is non-Newtonian. The evolution
of the lubricating front does not differ substantially from the purely
Newtonian case, but the evolution of the top fluid front differs
substantially. Our model solutions are found consistent with
laboratory experiments \citep{KumarZuriKoganEtAl2021JoFMLubricated},
particularly in predicting the time exponents of the front evolution
and the thickness fields.  Discrepancies that we find in the
intercepts predictions, particularly that of the lubricating fluid
front, suggest that additional physical mechanisms may contribute to
the front evolution, such as hydrofracturing or wall-slip along the
substrate. Exploration of these mechanisms will be the topic of future
studies.
Our results have implications to the understanding of spatiotemporal
distribution of lubrication networks beneath ice sheets and to
understanding their stability.

AG was partially supported by VATAT High-TEC fellowship for excellent
women in science.  This research was supported by the GERMAN-ISRAELI
FOUNDATION (grant No. I240430182015).  Declaration of Interests. The
authors report no conflict of interest.

\appendix

\section{Validation of the numerical code}
\label{sec:numerical scheme}

\subsection{Non-lubricated gravity currents}
\label{num.huppert}

\begin{figure}[h]
    \begin{overpic}[scale=0.45]{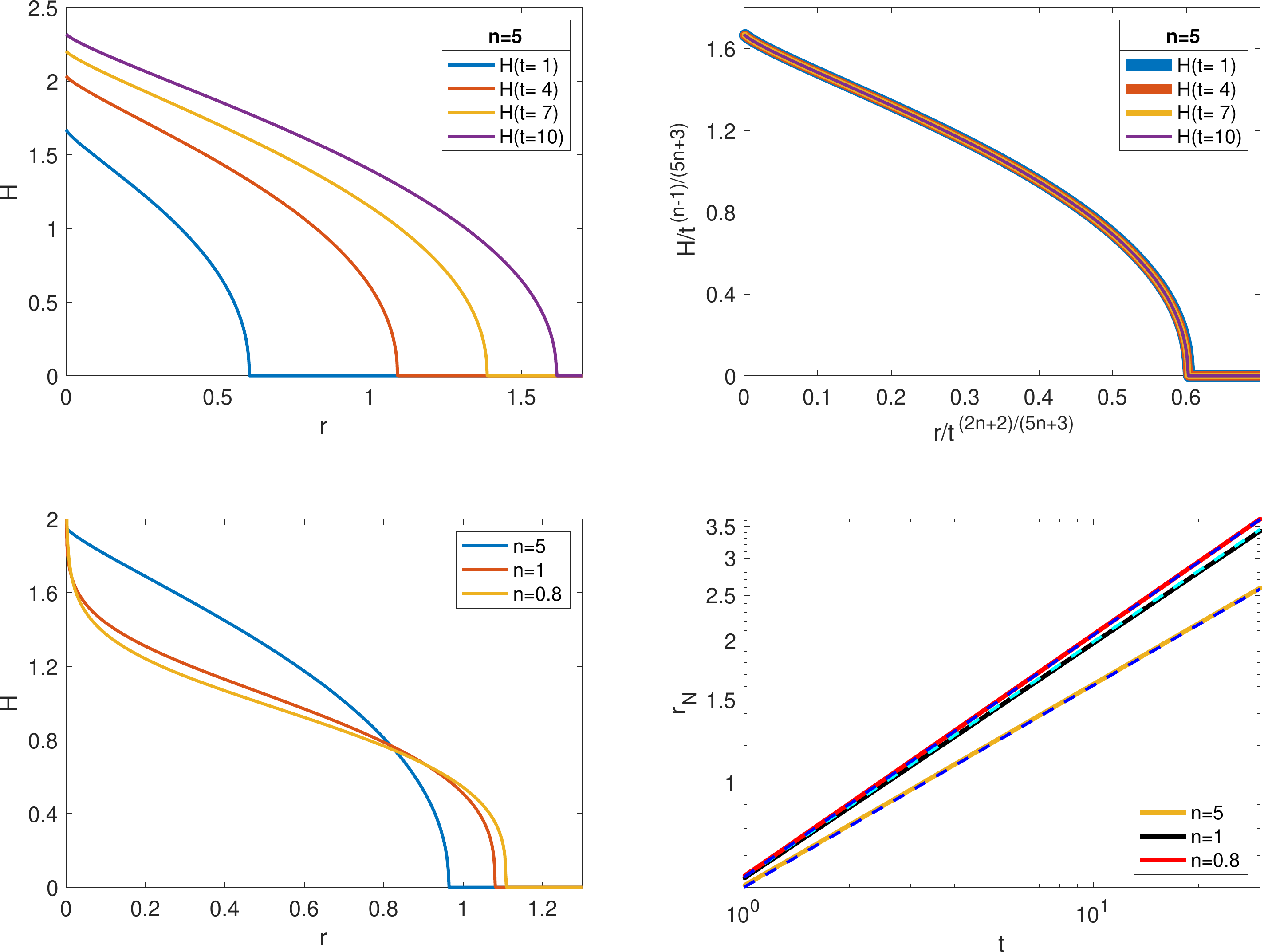}
     \put(58.7,19.5){\includegraphics[scale=0.089]{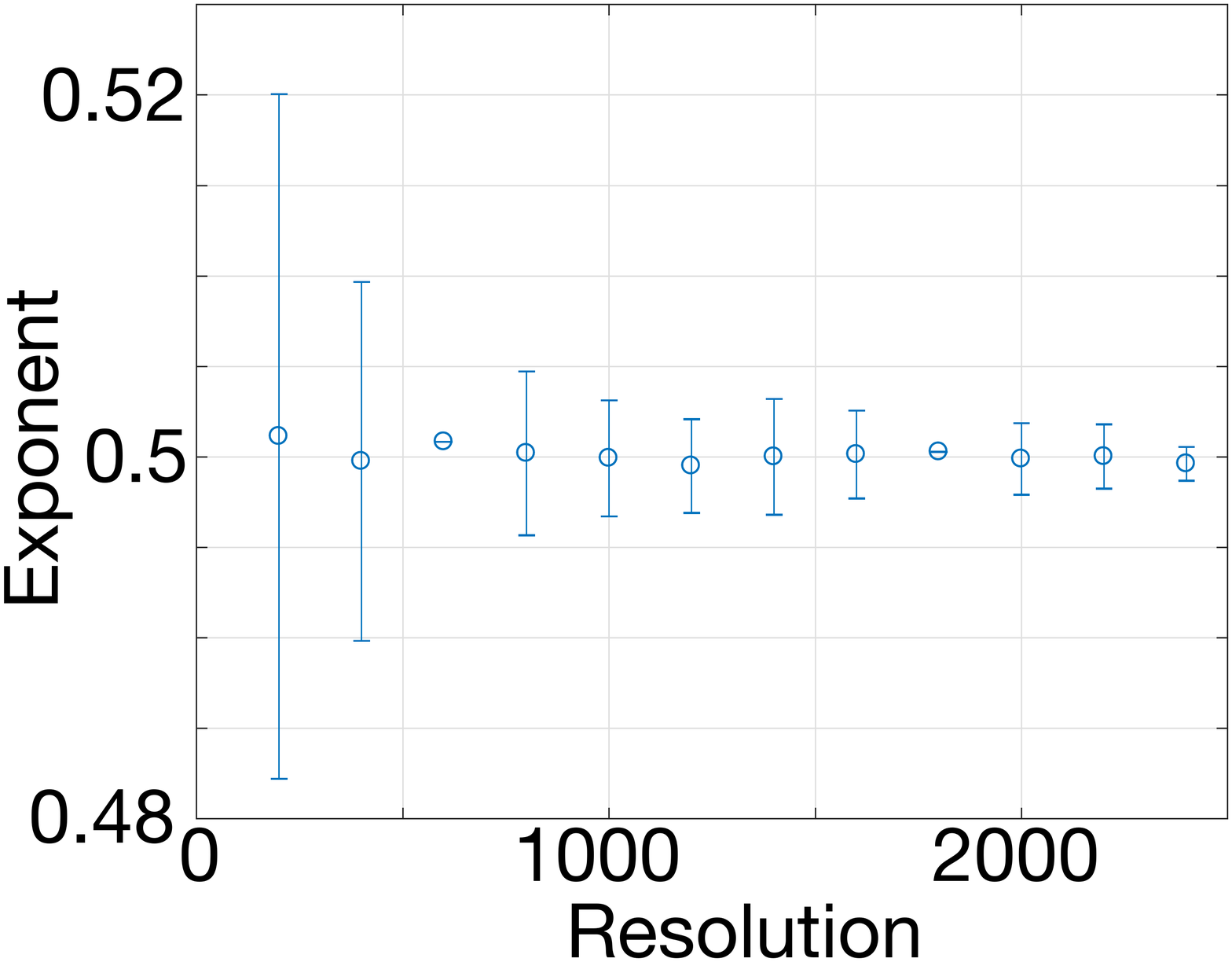}}   
   \end{overpic}
   \begin{picture}(0,0)(375,15)
     \put(5,163){(a)}
     \put(200,163){(c)} 
     \put(5,13){(d)}
     \put(200,13){(f)}
   \end{picture}
  \caption{Validation of the lubricated GC numerical solution with the
    theoretical prediction of non-lubricated GC of power-law fluids
    ($\Q=0, \a=1$) \citep{Sayag-Worster:2013-Axisymmetric}.  (a) Fluid
    height for n=5 at non-dimensional times t=1,4,7,10 in the regular
    thickness-radius space, and (b) in the thickness-radius space
    normalised by the theoretical prediction (c) Fluid height for n=5,
    1 and 0.8, at non-dimensional time t=3. (d) The front $r_N(t)$
    (solid), and the theoretical prediction for n=5, 1 and 0.8
    (dash). (inset) Regression results to the slope (exponent) of the
    numerical solution for $n=1$ as a function of spatial resolution
    in the range 200-2400 points in logarithmic spaced mesh.  Error
    bars represent the root-mean-square deviation of the fitted curve
    to the fronts.
    \label{SW0}}
\end{figure}

The model we develop in \S \ref{sec:model} describes in the limit
\(\mathscr{Q}=0\) a non-lubricated GC that propagates under no-slip
condition along the substrate. Such flow is similar to the flow in the
non-lubricated region, and is known to have a similarity solution
\citep{Sayag-Worster:2013-Axisymmetric}
\begin{equation}
  \label{eq:swsol}
h(r,t)\propto t^{\frac{n-1}{5n+3}},\qquad  r_N(t) \propto t^{\frac{2n+2}{5n+3}},
\end{equation}
for constant flux \(\a=1\).  We use this solution to validate our
numerical solution in the $\Q=0$ limit. Specifically, we solve the
dimensionless equation set (\S\ref{dimensionless}) with
\(\mathscr{Q}=0\), zero initial thickness \(H(r,0)=0\), and
logarithmically spaced spatial mesh with 1200 points. We find our
solutions for the fluid height and for the leading front consistent
with the theoretical predictions (Figure \ref{SW0}).  Repeating the
same computation for varying spatial resolutions, we find that the
convergence accuracy of the front exponent to the predicted
theoretical value grows with a the number of spatial grid points
(Figure \ref{SW0}d, inset).




\subsection{Lubricated Newtonian gravity currents}
\label{num.kw}

\begin{figure}[h]
\centering
\includegraphics[scale=0.6]{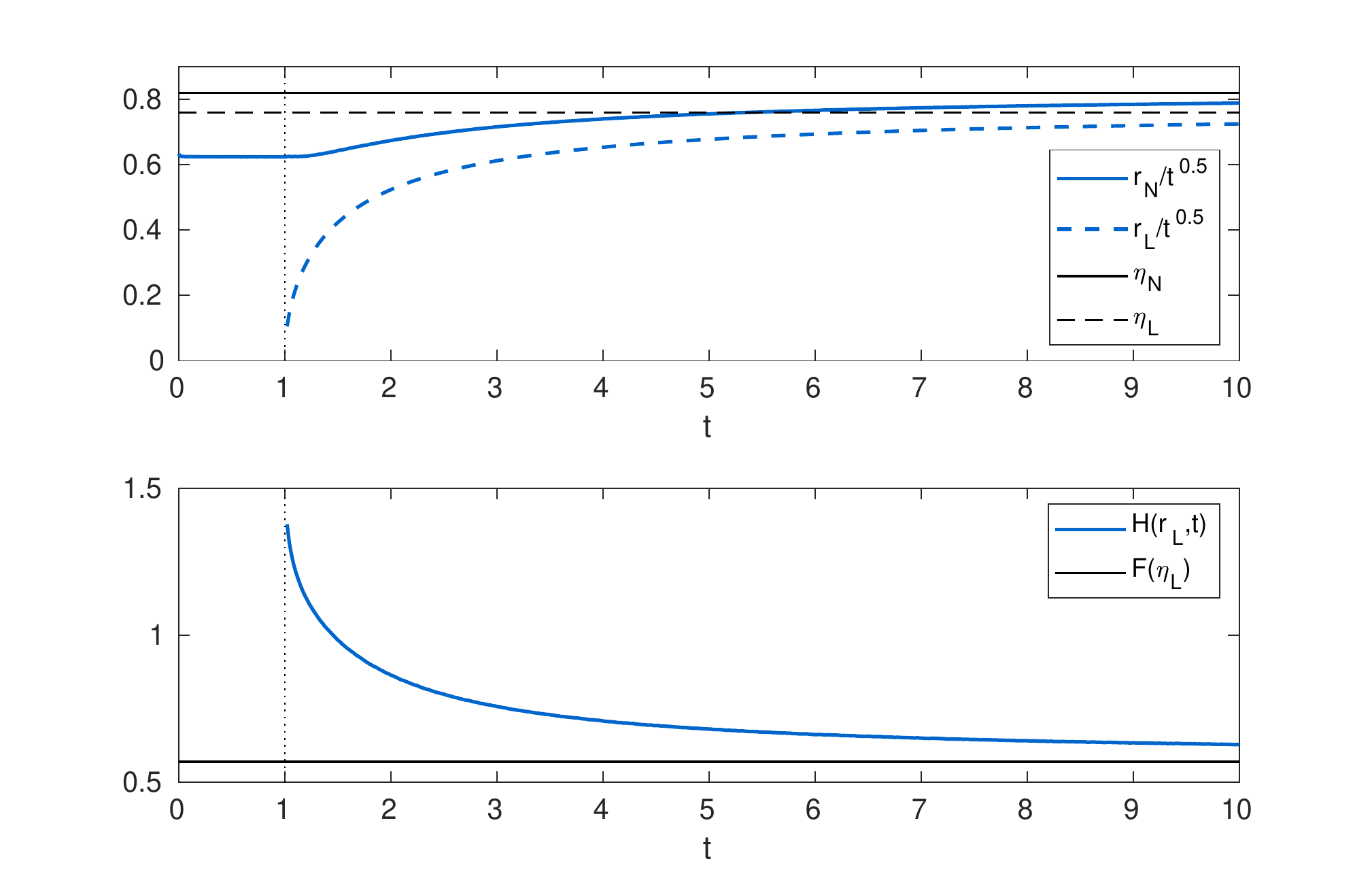}
\caption{Validation of the lubricated GC numerical
    solution with the Newtonian ($n=1$) lubricated GC
  (blue) with \(\mathscr{M}=10,000\), \(\mathscr{Q}=0.2\) and
  \(\mathscr{D}=0.1\), showing the convergence of the normalized fronts
  (top) and the lubricated fluid height at lubricating fluid front,
  \(r_L\) (bottom) to the theoretically predicted constant values
  (black).
  \label{logfig1}}
\end{figure}

In the limit $n=1, \a=1$ our numerical model converges to the purely
Newtonian lubricated GC discharged at constant flux
\citep{KowalWorster:2015-JFM-Lubricated}. 
Considering first the specific case where \(\mathscr{M}=10000\),
\(\mathscr{Q}=0.2\) and \(\mathscr{D}=0.1\), we find that both the
fronts $r_N,r_L$ and the upper fluid height at the lubricant front
$H(r_L)$ converge to the theoretical values $\eta _N,\eta _L$ and
\(F(\eta)\), respectively (Figure \ref{logfig1}).
Second, we find that our solutions for the coefficients \(\eta_N\) and
\(\eta_L\) is consistent with the theoretical predictions for a wide
range of \(\mathscr{M}\) values (Figures \ref{fig:betaNL_cNL}. As
shown in Appendix \ref{num.huppert}, small discrepancies from the
predicted values are due to low spatial resolution.
Lastly, the solutions to the specific regime discussed in
\S\ref{sec:const} (Figures \ref{figure3}) is another evidence for the
consistency between our numerical results and those of \cite[][Figure
13]{KowalWorster:2015-JFM-Lubricated}.

\bibliographystyle{jfm}
\bibliography{GS2020}

\end{document}